\documentclass[12pt]{article}
\usepackage{amsmath,amssymb,amsfonts,epsfig,graphicx,euscript}%
\usepackage[bookmarks,bookmarksnumbered,linktocpage,pdfstartview=FitH]{hyperref}
\hypersetup{colorlinks,%
citecolor=red,%
filecolor=blue,%
linkcolor=blue,%
urlcolor=blue,%
pdftex}
\usepackage{amsmath}
\usepackage{amsfonts}
\usepackage{graphicx}
\usepackage{caption}
\usepackage{subcaption}
\usepackage{float}
\usepackage[all]{hypcap}
\numberwithin{equation}{section}
\usepackage{cite}
\usepackage{calc}
\newlength{\spacer}
\newsavebox{\mybox}

\newcommand{\bse}{\begin{subequations}}
\newcommand{\ese}{\end{subequations}}
\newcommand{\be}{\begin{equation}}
\newcommand{\ee}{\end{equation}}
\newcommand{\bea}{\begin{eqnarray}}
\newcommand{\eea}{\end{eqnarray}}
\newcommand{\ba}{\begin{array}}
\newcommand{\ea}{\end{array}}

\begin{document}

\renewcommand*{\thefootnote}{\fnsymbol{footnote}} 

\begin{center}
{ \large{\textbf{A Minimal System Including Weak Sphalerons for Investigating the Evolution of Matter Asymmetries and Hypermagnetic Fields}}} 
\vspace*{0.5cm}
\begin{center}
{\bf  S. Rostam Zadeh\footnote{sh$_{-}$rostamzadeh@ipm.ir},$^1$  S. S. Gousheh\footnote{ss-gousheh@sbu.ac.ir}$^2$}\\%
\vspace*{0.5cm}
{\it {$^1$School of Particles and Accelerators, Institute for Research in Fundamental Sciences (IPM), P.O.Box 19395-5531, Tehran, Iran\\$^2$Department of Physics, Shahid Beheshti University, G.C., Evin, Tehran 19839, Iran}}  \\
\vspace*{0.5cm}
\end{center}
\end{center}

\renewcommand*{\thefootnote}{\arabic{footnote}} 
\setcounter{footnote}{0}

\begin{center}
\textbf{Abstract}
\end{center}

We study simultaneous evolution of large scale hypermagnetic fields and the asymmetries of quarks, leptons and Higgs boson, in the temperature range $100$GeV$\leq T \leq 10$TeV. Above $10$TeV, we identify all of the major fast interactions and use the associated conservation laws as constraints on the initial conditions at $10$TeV. Below $10$TeV, we identify the major processes which fall out of equilibrium or emerge as non-negligible processes and derive the relevant evolution equations. These include the Abelian anomalies which violate fermion numbers, direct and inverse Higgs decays that change the chiralities of fermions, and weak sphalerons which violate the left-handed fermion numbers. We also consider the contributions of all fermionic chemical potentials to the U$_\textrm{Y}$(1) Chern-Simons term which affects the evolution through the AMHD equations. Thus, we present a minimal set of self-consistent initial conditions and evolution equations, which respect all constraints coming from conservation laws, fast processes and charge neutrality of the plasma. We solve the coupled evolution equations and find that initial large hypermagnetic field can produce matter asymmetries starting from zero initial value, and vice versa provided an initial seed of hypermagnetic field is present and the rate of the electron Yukawa processes is lower. We find that our model yields acceptable values for baryon asymmetry and magnetic field. However, the scale of the magnetic field obtained is much smaller than the observational data, even when the turbulence driven inverse cascade mechanism in the broken phase is taken into account.

\newpage

\tableofcontents

\section{Introduction}\label{Introduction}
The origin of the matter anti-matter asymmetry of the Universe is one of the great problems in cosmology. It is believed that, as the temperature in the early Universe decreased, nearly all of the particles and anti-particles annihilated one another just before the hadronization phase, and a small amount of matter remained to be the source for the matter in the present day Universe \cite{Canetti}. Two independent sources of information, namely the abundances of light elements in the intergalactic medium (IGM) \cite{Fields}, and the power spectrum of the temperature fluctuations in the cosmic microwave background (CMB) \cite{Simha} determine the value of the baryon asymmetry as $\eta_B\sim6\times 10^{-10}$. As Sakharov suggested, three necessary conditions are needed for the dynamical creation of this asymmetry from an initial state which is matter anti-matter symmetric. They are: the existence of baryon number violation processes, C and CP violation \cite{Wu,Christenson}, and deviation from thermal equilibrium \cite{Sakharov}. 

Another great problem facing the cosmology is the origin of the long-range magnetic fields detected in some galaxies \cite{Kronberg2008,Bernet,Wolfe}, galaxy clusters \cite{Clarke, Bonafede, Feretti} and high redshift protogalactic structures \cite{Kandus}. It is widely believed that these magnetic fields are generated from the amplification of some initial seed fields \cite{Harrison}, whose nature is largely unknown \cite{Kronberg, Kulsrud}. The extensive presence of the magnetic fields at high redshifts, as well as the presence of the coherent magnetic fields in the intergalactic medium \cite{Vovk,Neronov,Tavecchio,Tavecchio2011,Ando,Essey}, strengthens the idea of primordial magnetism \cite{Kandus}. Therefore, our universe in its hot early stages might have contained some magnetic fields.

At high temperatures, the non-Abelian gauge fields acquire a magnetic mass gap $\sim$ $g^2T$ \cite{Gross}, while the Abelian one remains massless \cite{Kajantie}. As a result, the Abelian U(1) magnetic field is the only long-range magnetic field surviving in the plasma. In the symmetric phase, the chiral coupling of the U$_\textrm{Y}$(1) gauge fields to the fermions leads to the fermion number violation. The anomalous coupling of the hypercharge fields to fermion number densities shows up both in the Abelian anomaly equations of the form $\partial_\mu j^\mu \sim \frac{g'^2}{4\pi^2} \textbf{E}_\textbf{Y}.\textbf{B}_\textbf{Y}$, and in the U$_\textrm{Y}$(1) Chern-Simons term. This term is induced in the effective Lagrangian density of the U$_\textrm{Y}$(1) gauge field and gives rise to an anomalous term in the magnetohydrodynamic (MHD) equations \cite{Laine,shiva,Giovannini,Joyce}.

The $\textrm{SU}(2)_{\textrm{L}}$ gauge fields couple to the fermions chirally as well. This leads to the emergence of the $\textrm{SU}(2)_{\textrm{L}}$ Chern-Simons term in the effective Lagrangian density of the corresponding gauge fields \cite{Redlich,Rubakov,Laine,shiva}, and the existence of the $\textrm{SU}(2)_{\textrm{L}}$ anomaly equations (see Appendix B of Ref. \cite{Long}). The non-perturbative high-temperature effects associated with the $\textrm{SU}(2)_{\textrm{L}}$ non-Abelian anomaly, known as the weak sphalerons, are widely investigated in the literature \cite{Kuzmin}. They actively participate in most of the matter asymmetry generation scenarios. Indeed, they change the baryon and the lepton asymmetries ($\eta_B$ and $\eta_L$) simultaneously via the violation of left-handed quark and lepton numbers, while respecting the conservation of $\eta_B-\eta_L$ \cite{Shaposhnikov}. It has been argued that in the early Universe and in the absence of the hypermagnetic fields, the weak sphalerons can wash out the baryon asymmetry of the Universe unless it is encoded in a $\eta_B-\eta_L$ asymmetry \cite{Gorbunov}.
As mentioned earlier, the weak sphalerons act only on the left-handed fermions; therefore, the washout process is completed when the Yukawa interactions and the weak interactions are also in thermal equilibrium \cite{Harvey}.


Right-handed electrons play a crucial role in some of the suggested scenarios in cosmology \cite{Laine,Dvornikov,Giovannini,Dvornikov2011,Campbell,Joyce,shiva,Smirnov,Sokoloff,shiva2}.
Indeed, for $T>T_{RL} \simeq 10$ TeV, they are decoupled from the thermal ensemble and their number density is conserved
\cite{Campbell}.
\ This is due to the fact that, the Yukawa coupling of the electrons with the Higgs bosons $h_e$ is tiny. 
Therefore, the electron chirality flip processes\footnote{In addition to the direct and inverse Higgs decays, some gauge and fermion scattering processes (such as $e_R H\leftrightarrow L_e A$, where $A=Y$ or $W$, and $e_R L_f\leftrightarrow L_e f_R$) contribute to the chiralty flip rate of electrons as well (see the third paper of Ref.\ \cite{Campbell}).} 
 (e.g. direct and inverse Higgs decays in reactions $e_L \bar{e}_R\leftrightarrow\phi^{(0)}$ and $\nu_e^L \bar{e}_R\leftrightarrow\phi^{(+)}$, and their conjugate reactions) whose rates $\sim h_e^2T $ are much lower than the Hubble expansion rate, are out of thermal equilibrium in this range of temperatures \cite{Campbell}.\footnote{For $T>T_{RL}$, the number density of right-handed electrons is conserved even if the Abelian anomaly ($\partial_\mu j_{e_R}^\mu = \frac{g'^2}{4\pi^2} \textbf{E}_\textbf{Y}.\textbf{B}_\textbf{Y}$) is taken into account. Indeed, as our studies show, the abelian anomalous effects are strong near the electroweak phase transition (EWPT).}

Using the above fact, the authors of \cite{Campbell} suggested the possibility to encode the baryon asymmetry in a right-handed electron asymmetry protected from the weak sphalerons down to $T_{RL}$.\footnote{$T_{RL}$ as computed in the first paper of Ref.\ \cite{Campbell} was $\sim 1$TeV.} They argued that, at temperatures below $T_{RL}$, the electron chirality flip processes come into equilibrium, while the weak sphalerons start to fall out of equilibrium.\footnote{In recent years, $T_{sph}$ at which the weak sphaleron processes fall out of thermal equilibrium is computed as $\sim 135\ $GeV \cite{Burnier}.} Therefore,  
they may not be able to turn the generated left-handed leptons into antiquarks to erase the remnant baryon and lepton asymmetries. This would raise the possibility to preserve an initial baryon asymmetry when $\eta_B-\eta_L=0$. However, the relevant studies showed the failure of this scenario to preserve the asymmetries against the weak sphalerons in the absence of the hypermagnetic fields \cite{Campbell}.


The main purpose of this paper is to build a minimal model to study the simultaneous evolution of the matter asymmetries and the hypermagnetic fields in the temperature range, $T_{EW}\simeq 100 \textrm{GeV}<T<T_{RL} \simeq 10 \textrm{TeV}$, which takes into account the most important processes; and more importantly, constitutes a minimal set of self-consistent assumptions, initial conditions and evolution equations which respect all constraints coming from the equilibrium conditions of fast processes, the conservation laws in the Standard Model and the charge neutrality of the plasma. 
To accomplish this task, we identify all of the major fast processes and conservation laws above $10$TeV and use them as constraints on the initial conditions at $T=10$TeV. Then we identify the major processes that should be taken into account for $T<10$TeV and derive the evolution equations, respecting the remaining conservation laws. The latter processes include the weak sphalerons which affect the evolution equations strongly due to their high rate in the symmetric phase \cite{Burnier}. 
We consider the quark and the lepton asymmetries of all generations and include the contributions of their chemical potentials to the U$_\textrm{Y}$(1) Chern-Simons term.
We assume nonzero Higgs asymmetry and consider direct and inverse Higgs decay processes operating on the quarks and the leptons. The former is necessary for maintaining constraints such as charge neutrality of the plasma.
\footnote{The assumption of zero Higgs asymmetry is an extra constraint which takes the place of one of the main constraints of plasma such as the charge neutrality condition.}   
For temperatures below $T_{RL}$, we reduce the number of dynamical equations by using the conservation laws for the hypercharge, $\eta_B/3-\eta_1$, $\eta_B/3-\eta_2$, $\eta_B/3-\eta_3$,\footnote{These asymmetries are defined below Eqs.\ (\ref{charge2}).} and considering simplifications for the asymmetries of the quarks and the tau lepton due to the fast gauge and Yukawa processes acting on them. We also use these simplifications for the coefficient of the U$_\textrm{Y}$(1) Chern-Simons term.

In this work, we focus on models with vanishing $\eta_B-\eta_L$. We assume the presence of the weak sphalerons and address the question of whether the observed baryon asymmetry of the Universe can arise entirely from the decaying magnetic helicity of a primordial magnetic field (i.e., BAU-from-PMF). We also address the question of whether it is possible for a tiny seed of the hypermagnetic field to grow in the presence of initial matter asymmetries and weak sphalerons (i.e., PMF-from-BAU). We investigate both scenarios in the interesting spots of the parameter space. Moreover, since the electron asymmetries do play a key role, many previous studies have focused on simply the electron asymmetries. A few studies have included kinetic equations for all of the Standard Model fermion species (see Ref.\ \cite{Kamada}). Here, we also include the effects of all fermionic asymmetries and Higgs asymmetry in our model. However, the main advantage of our model is that we use the constraints coming from the conservation laws and fast processes in the electroweak plasma to significantly reduce the number of necessary and consistent kinetic equations to three, then simply obtain all other asymmeties in terms of them. Most importantly, we use the constraints coming from the conservation laws and fast processes to obtain a consistent set of initial values for all asymmetries which can be calculated by fixing the right-handed electron asymmetry. Indeed, only those initial matter asymmetries are valid which satisfy all of the aforementioned constraints. Most of the previous studies have not emphasized the necessity of consistency of the initial conditions with these constraints, whose neglect can lead to the violation of some conservation laws such as charge neutrality of the plasma. We also include the chiral magnetic effect (CME) in our model through the Abelian Chern-Simons term and observe that the CME suppresses the growth of the baryon asymmetry in both scenarios. Furthermore, in the PMF-from-BAU scenario, this term is crucial for the growth of the hypermagnetic field and in its absence no strengthening happens for the magnetic field (see also Ref.\ \cite{shiva2}). We also observe that in the absence of this term the results become sensitive to the chirality flip rates. The CME is an important effect that has not been taken into account in some previous studies. We also use the Yukawa rates which are estimated in Ref.\ \cite{Kamada}. Since, the rate of the electron Yukawa processes is an important parameter with a key role and there is an uncertainty in it, we investigate the effect of changing this rate on the results in both of our scenarios via multiplying it with an adjustable parameter.

The outline of our paper is the following. In Section \ref{Equilibrium Conditions}, we present the equilibrium conditions and conservation laws governing the system for temperatures above $T_{RL}$. In Section \ref{Nonequilibrium Phase}, we derive the evolution equations of the asymmetries and the hypermagnetic field.
In Subsection \ref{Nonequilibrium and Almost Equilibrium Processes}, we categorize all of the relevant processes according to their rates above and below $T_{RL}$. In Subsection \ref{Static Hypermagnetic Chern-Simons Term}, we start with the general form of the U$_\textrm{Y}$(1) Chern-Simons term \cite{shiva}, then rewrite it in a new form and simplify it.    
In Subsections \ref{The Dynamical Equation of the Hypermagnetic Field} and \ref{The Evolution Equations of the Lepton and Baryon Asymmetries}, we derive the required dynamical equations for the asymmetries and the hypermagnetic field, considering the Abelian anomaly, the weak sphalerons, the chirality flip processes through direct and inverse Higgs decays, and the simplified coefficient of the hypermagnetic Chern-Simons term obtained in Subsection \ref{Static Hypermagnetic Chern-Simons Term}.
In Section \ref{Results}, we numerically solve the set of coupled differential equations for the asymmetries and the hypermagnetic field 
for some interesting ranges of initial conditions and present the results. 
We use the conventions stated in Appendix A and the anomaly equations summarized in Appendix B of Ref. \cite{Long}. We also use the derivation method of the kinetic equations for the lepton asymmetries given in Appendix B of Ref. \cite{Dvornikov}. In Section \ref{Summary and Discussion} we summarize the main results and state our conclusions.

\section{Equilibrium Conditions}\label{Equilibrium Conditions}
In this section we consider the equilibrium conditions established by the fast processes, i.e.\ the ones whose rates are much higher than the Hubble expansion rate at temperatures above $T_{RL}$.\footnote{The results given in Section \ref{Results} show that strong hypermagnetic fields have the ability to make some of the reactions fall out of chemical equilibrium, especially near the electroweak phase transition (EWPT). 
However, for temperatures above $T_{RL}$, the term corresponding to the hypermagnetic fields in the evolution equations is negligible, and therefore the usual assumption of chemical equilibrium for the fast reactions still remains valid, at least for the values of the hypermagnetic field amplitude assumed in this work.} 

Since the non-Abelian gauge interactions are in thermal equilibrium at all temperatures of concern, they force to equalize the asymmetries of different components of all multiplets \cite{Long}. So, let us denote the common chemical potential of left-handed (right-handed) leptons by $\mu_{L_i}$($\mu_{R_i}$), the left-handed quarks with different colors by $\mu_{Q_i}$, and up (down) right-handed quarks with different colors by $\mu_{{u_R}_i}$ ($\mu_{{d_R}_i}$), where \lq{\textit{i}}\rq\ is the generation index.  

There are also other fast processes operating on the quarks, and the second and third generation leptons. 
They are: up-type Yukawa in processes $u_R^i\bar{d}_L^i\leftrightarrow\phi^{(+)}$ and $u_R^i\bar{u}_L^i\leftrightarrow\phi^{(0)}$, down-type Yukawa in processes $d_R^j\bar{u}_L^i \leftrightarrow \phi^{(-)}$ and $d_R^j\bar{d}_L^i \leftrightarrow \tilde{\phi}^{(0)}$, electron-type Yukawa in processes $e_R^i\bar{\nu}_L^i \leftrightarrow \phi^{(-)}$ and $e_R^i\bar{e}_L^i \leftrightarrow \tilde{\phi}^{(0)}$, and their conjugate reactions \cite{Long}.\footnote{See Section 2.3.2 of Ref.\ \cite{Kamada} for a more complete list of Yukawa reactions and also Appendix B of Ref.\ \cite{Fujita} for the values of the corresponding Yukawa couplings.} Assuming that these Yukawa interactions for the aforementioned particles are in equilibrium, we obtain
\be\begin{split}\label{Yukawa1}
\mu_{{u_R}_i}-\mu_{Q_i}=\mu_0,\ \ \ \ \ \ \ \ \ \ \ \ \ \ \ \ \ \ \ \ \ \ \ \ \ \ \ \ \ \ \ \ i=1,2,3,\cr
\mu_{{d_R}_j}-\mu_{Q_i}=-\mu_0,\ \ \ \ \ \ \ \ \ \ \ \ \ \ \ \ \ \ \ \ \ \ \ \ \ \ \ i,j=1,2,3,\cr
\mu_{R_i}-\mu_{L_i}=-\mu_0;\ \ \ \ \ \ \ \ \ \ \ \ \ \ \ \ \ \ \ \ \ \ \ \ \ \ \ \ \ \ i=2,3,\ \ \    
\end{split}\ee
where $\mu_0$ is the chemical potential of the Higgs field. 
As a result of the flavor mixing in the quark sector, all up or down quarks belonging to different generations with distinct handedness have the same chemical potential, i.e., $\mu_{{u_R}_i}=\mu_{u_R},\ \mu_{{d_R}_i}=\mu_{d_R},\ \textrm{and}\ \mu_{Q_i}=\mu_Q$, where $i=1,2,3$.\footnote{See Section 3 of the third paper of Ref.\ \cite{Campbell}.} Then, we obtain
\footnote{The strong sphaleron processes are also taken into account which lead to the constraint $\mu_{u_R}+ \mu_{d_R}=2\mu_Q$ (See Table 1 of Ref.\ \cite{Long}). However, this is not a new equation since it can be obtained from Eqs.\ (\ref{Yukawa}).} 
\be\begin{split}\label{Yukawa}
\mu_{u_R}-\mu_Q=\mu_0,\ \ \ \ \ \ \mu_{d_R}-\mu_Q=-\mu_0.
\end{split}\ee
Furthermore, using the above relations, the whole baryonic chemical potential simplifies as shown below 
\be\label{mu_B} 
\mu_B = \frac{1}{N_c}{\sum}_{i=1}^{n_G}\left[N_cN_w\mu_{Q_i} +N_c\mu_{u_{R_i}} +N_c\mu_{d_{R_i}}\right] = 12\mu_Q, 
\ee
where $N_c = 3$ and $N_w = 2$ are the ranks of non-Abelian gauge groups and $n_G=3$ is the number of generations.

Since the weak sphaleron processes are in equilibrium, they impose a further condition on the chemical potentials of left-handed quarks and leptons of all generations,\footnote{See Table 1 of Ref.\ \cite{Long}.} 
\be\label{sphaleron}
c_E=9\mu_Q +\mu_{L_1}+\mu_{L_2}+\mu_{L_3}=0.
\ee
Using the relations between the chemical potentials given by Eqs.\ (\ref{Yukawa1},\ref{Yukawa}), the equation for the charge neutrality of the electroweak plasma reduces to
\be\label{charge}
Q=6\mu_Q-\mu_{R_1}-\mu_{L_1}-2\mu_{L_2}-2\mu_{L_3}+13\mu_0=0.
\ee
In addition to Eqs.\ (\ref{sphaleron}) and (\ref{charge}) which include six unknown parameters, there are also four additional constraints on the chemical potentials so that the system has a unique solution. These conditions are expressed via the following conservation laws\footnote{We are working within the context of the Standard Model where the neutrinos are considered massless. The tiny masses of neutrinos in the broken phase point to a corresponding small mixing in the lepton sector which can be taken into account.}
\be\begin{split}\label{charge2}
\frac{\eta_B}{3}-\eta_1=c_1,\cr
\frac{\eta_B}{3}-\eta_2=c_2,\cr
\frac{\eta_B}{3}-\eta_3=c_3,\cr
\eta_{R_1}=c_{R_1},
\end{split}\ee
where $\eta_B=12\eta_Q$ is the whole baryon asymmetry, $\eta_i = 2\eta_{L_i} + \eta_{R_i}$ is the lepton asymmetry of the \textit{i}th-generation, and the constants $c_1,\ c_2,\ c_3$ and $c_{R_1}$ are the primordial values. Thus, the initial conditions are uniquely determined by specifying the values of $c_1,\ c_2,\ c_3$ and $c_{R_1}$. The relation between the matter asymmetry $\eta$ and the chemical potential $\mu$ is \cite{Gorbunov}
\be\label{asym-chem}
\eta \equiv \frac{n-{\bar{n}}}{s} \simeq \frac{\mu T^2 c}{6s} + O((\frac{\mu}{T})^3) = \frac{15c}{4\pi^2g^*} \frac{\mu}{T} + O((\frac{\mu}{T})^3),
\ee
where $c$ is 1 for the fermions and 2 for the bosons, $s = \frac{2\pi^2}{45} g^*T^3$ is the entropy density of the Universe, and $g^*=106.75$ is the effective number of relativistic degrees of freedom. In this paper, we assume that there is no primordial asymmetry for $\eta_B/3-\eta_i$, namely $c_1=c_2=c_3=0$. However, there might be a primordial asymmetry for right-handed electrons reflected in a nonzero value for $c_{R_1}$. We solve the six mentioned equations with the above assumptions to obtain the initial values of the asymmetries at $T=T_{RL}$.

\section{Dynamical Phase}\label{Nonequilibrium Phase}
In Section \ref{Equilibrium Conditions}, we discussed the equilibrium conditions in the electroweak plasma when the temperature is above $T_{RL}$. In this section, we investigate the dynamical phase in the temperature range $T_{EW}<T<T_{RL}$. 
Taking into account the relevant conservation laws and assuming that the processes operating on the quarks and the tau lepton\footnote{The processes for the muon are marginally fast. Hence, we include the evolution equation for the muon. 
} are still nearly in equilibrium, we obtain the minimum number of required dynamical equations and simplify them. We then study the evolution of matter asymmetries and long-range hypermagnetic fields by solving these evolution equations numerically.

\subsection{Categorization of the Relevant Processes}\label{Nonequilibrium and Almost Equilibrium Processes}

Let us now investigate the effects of the most important processes in the temperature range $T_{EW}<T<T_{RL}$. To do this, we find it useful to label each of the processes according to its rate as either ``Fast", ``Dynamical" or ``Slow". Fast processes are the ones that are nearly in equilibrium in the whole temperature range under consideration. There are usually constant or conserved quantities associated with these processes. Slow processes have rates much smaller than the Hubble expansion rate and are not only out of equilibrium, but also are almost inactive or frozen in the temperature range under consideration. Dynamical processes have intermediate rates in the temperature range under consideration and their dynamics is interesting for us.\ 
Now we can divide these processes into three categories according to their rate for $T>T_{RL}$ and for $T<T_{RL}$, and tabulate them according to these categories in Table 1. The first category is denoted by ``Fast-Fast" indicating the processes that are fast both for $T>T_{RL}$ and for $T<T_{RL}$. The second category is denoted by ``Fast-Dynamical" indicating the processes that are fast for $T>T_{RL}$ and dynamical for $T<T_{RL}$.\footnote{The weak sphaleron processes can be categorized as Fast processes (with $c_E=0$) for $T<T_{RL}$. However, we categorize them as Dynamical in order to investigate the dynamics of them as an interesting representative of Fast processes. Therefore, we do not force $c_E$ to stay at zero and let it evolve according to the evolution of its constituents.} The third category is denoted by ``Slow-Dynamical" indicating the processes that are slow for $T>T_{RL}$ and dynamical for $T<T_{RL}$.\footnote{The Abelian anomalous effects appearing in the Abelian anomaly and U$_\textrm{Y}$(1) Chern-Simons terms are weak for $T>T_{RL}$ and gradually become strong for $T<T_{RL}$. Therefore, they are also categorized as Slow-Dynamical.} 

\begin{table}[ht]
\caption{Major processes categorized according to their behavior for $T>T_{RL}$ and $T<T_{RL}$, respectively.}
\label{table:processes}
\centering
\begin{tabular}{|c|c|}
\hline
Fast-Fast & Gauge Interactions, quark Chirality flip, $\tau$ Chirality flip \\[1ex]
\hline
Fast-Dynamical & Weak Sphalerons, $\mu$ Chirality flip \\[1ex]
\hline
Slow-Dynamical & Abelian anomaly, U$_\textrm{Y}$(1) Chern-Simons, $e$ Chirality flip \\[1ex]
\hline
\end{tabular}
\end{table}

In this table the ``Chirality flip" processes refer to the ones which are driven by the Higgs.\footnote{Fast strong sphaleron processes that change the chiralities of the quarks are also considered. However, as mentioned in Footnote 10, they do not lead to any new constraint.} The fast processes for $T>T_{RL}$ are used to impose constraints on the initial conditions at $T=T_{RL}$, and for $T<T_{RL}$ to impose constraints for reducing the number of dynamical equations. We now qualitatively discuss these processes, and a detailed quantitative analysis starts in the next subsection. As the temperature decreases, passing through $T_{RL}$ and moving towards $T_{EW}$,   the U$_\textrm{Y}$(1) Chern-Simons term and the Abelian anomaly term in the evolution equations gradually gain strength. These terms are connected with the strength of the hypermagnetic field and, as we shall show, they generically increase the amplitude of this field.\footnote{In this study, either the hypermagnetic field is strong from the beginning or it becomes strong due to the initial matter asymmetries.} Subsequently, strong hypermagnetic field becomes able to make some of the fast processes, {\it i.e.} the ones that are in competition with it in the evolution equations, fall out of chemical equilibrium. These processes include the Higgs driven chirality flip and the weak sphaleron processes. This is in spite of the fact that, as the temperature decreases, generically both the Higgs driven chirality flip processes and the weak sphaleron processes gain strength as compared to the Hubble expansion rate. Obviously, the larger the rates of these processes, the less they go out of equilibrium.\footnote{It can be seen in Figure \ref{3.45ys} that the electron chirality flip reactions which have low rates, fall out of chemical equilibrium more, as compared to the muon chirality flip processes and the weak sphalerons whose rates are much higher.} As a result, the equilibrium conditions for the quarks and the tau lepton given by Eqs.\ (\ref{Yukawa1},\ref{Yukawa}) still remain valid to a good approximation, at least for the range of the hypermagnetic field amplitudes that we consider in this paper. 
However, that of the muon, as given by Eq.\ (\ref{Yukawa1}), does not remain valid in the whole interval under consideration especially near the EWPT (see Figures \ref{3.45ys} and \ref{21-}). The equilibrium condition for the weak sphaleron processes given by Eq.\ (\ref{sphaleron}) remains valid to a good approximation as well. However, as mentioned in Footnote 14, we assume that the constraint imposed by the weak sphaleron processes is no longer valid, in order to be able to investigate the dynamics of these fast processes.
\footnote{In the temperature region under consideration, i.e. $T_{EW}<T<T_{RL}$, there are subregions where the weak sphaleron processes and the muon Yukawa reactions are in equilibrium. No inconsistency occurs when we discard the two equilibrium conditions and the ensuing constraints in the whole region, since we let $c_E=9\mu_Q +\mu_{L_1}+\mu_{L_2}+\mu_{L_3}$ and $c_{\mu}=\mu_{R_2}-\mu_{L_2}+\mu_0$ evolve freely in accordance to the evolution of their constituents, when we solve the dynamical equations with various initial conditions. Then in the aforementioned subregions $c_E$ and $c_{\mu}$ attain constant values as a result of their evolution equations.}
Using Eqs.\ (\ref{Yukawa1},\ref{Yukawa}) for the chemical potentials of the quarks and the tau lepton, the equation for the charge neutrality of the electroweak plasma reduces to
\be\label{charge3}
Q=6\mu_Q-\mu_{R_1}-\mu_{L_1}-\mu_{R_2}-\mu_{L_2}-2\mu_{L_3}+12\mu_0=0.
\ee
There are also four conservation laws given by Eqs.\ (\ref{charge2}) for $T \geq T_{RL}$. The first three of these equations are always valid in the Standard Model, but the fourth one which is the conservation of the right-handed electron asymmetry is not valid any more, since the electron chirality flip processes become active below $T_{RL}$ (see Table \ref{table:processes}). Therefore, we have four constraints given by Eq.\ (\ref{charge3}) and the first three of Eqs.\ (\ref{charge2}) which include seven unknown parameters. 
So, we need three evolution equations for three of the asymmetries in addition to these four constraints to obtain the values of all asymmetries as a function of time. We derive the dynamical equations for the asymmetries of right-handed electron, left-handed electron and right-handed muon in subsection \ref{The Evolution Equations of the Lepton and Baryon Asymmetries}. The other asymmetries are obtained in terms of these three asymmetries by solving four constraint equations with the mentioned assumption of $c_1=c_2=c_3=0$, in the following form 
\be\begin{split}\label{other asymmetries}
\mu_Q=\frac{\mu_{R_1}+2\mu_{L_1}}{4},\ \ \ \ \ \ \ \ \ \ \ \ \ \ \ \ \ \ \mu_{L_2}=\frac{\mu_{R_1}+2\mu_{L_1}-\mu_{R_2}}{2},\ \ \cr
\mu_{L_3}=\frac{46\mu_{L_1}+24\mu_{R_1}+\mu_{R_2}}{68},\ \ \ \ \ \ \mu_0=\frac{4\mu_{R_1}+2\mu_{L_1}+3\mu_{R_2}}{68}.
\end{split}\ee 
In other words, obtaining the evolution of the asymmetries of left-handed electron, and right-handed electron and muon, one can use the above equations to obtain the evolution of the asymmetries of Higgs boson, and left-handed quarks, muon and tau lepton. Then, using Eqs.\ (\ref{Yukawa1},\ref{Yukawa}) one obtains the evolution of the asymmetries of right-handed up and down quarks and tau lepton as well.\footnote{We could as well have chosen to write the kinetic equation for the asymmetry of the left-handed muon instead of the right-handed one and to obtain other asymmetries in terms of the asymmetries of right-handed and left-handed electron and left-handed muon. The special role of right-handed and left-handed muons is their participation in the muon Yukawa prosess which is the most dynamical Yukawa process after the electron one. Ignoring this dynamics and considering the muon Yukawa process in equlibrium can change the values of all asymmetries and hypermagnetic field amplitude slightly.}

\subsection{Static U$_\textrm{Y}$(1) Chern-Simons Term}\label{Static Hypermagnetic Chern-Simons Term}
In the presence of fermionic chemical potentials and the hypermagnetic field, the U$_\textrm{Y}$(1) Chern-Simons term appears in the effective action of the U$_\textrm{Y}$(1) gauge field. In the static limit, one can derive the effective action for the soft gauge fields by implementing the method of dimensional reduction \cite{Ginsparg,Kajantie1995}. The Chern-Simons term in the Minkowskian effective Lagrangian density
is $-c'_E n'_{CS}$, where $n'_{CS}$ and $c'_E$ have the following forms \cite{Laine,shiva}
\be\begin{split}\label{c'_E}
n'_{CS} = \frac{g'^2}{32\pi^2}(2\textbf{Y}.\textbf{B}_\textbf{Y}),\ \ \ \ \ \ \ \ \ \ \ \ \ \ \ \ \ \ \ \ \ \ \ \ \ \ \ \ \ \ \ \ \ \ \ \ \ \ \ \ \ \cr
c'_E = {\sum}_{i=1}^{n_G}\left[-2\mu_{R_i} + \mu_{L_i} - \frac{2}{3}\mu_{{d_R}_i} - \frac{8}{3}\mu_{{u_R}_i} + \frac{1}{3}\mu_{Q_i}\right].
\end{split}\ee
In the above expressions, $n_G$ denotes the number of generations, $g'$ is the U$_\textrm{Y}$(1) gauge coupling, $\textbf{Y}$ is its corresponding vector potential, and $\textbf{B}_\textbf{Y}$ = $\nabla\times\textbf{Y}$ is the hypermagnetic field. We can rewrite $c'_E$ in the following form which enables us to simplify it
\be\begin{split}\label{c'_E_new}
c'_E = -2{\sum}_{i=1}^{n_G}\{\frac{1}{3}[(\mu_{{d_R}_i} -\mu_{Q_i}+\mu_0)+4(\mu_{{u_R}_i}-\mu_{Q_i}-\mu_0)]\cr
+(\mu_{R_i} - \mu_{L_i}+\mu_0)\}-{\sum}_{i=1}^{n_G}(3\mu_{Q_i}+\mu_{L_i}).
\end{split}\ee
Using the equilibrium expressions for the quarks and the tau lepton as given by Eqs.\ (\ref{Yukawa1},\ref{Yukawa}), Eq.\ (\ref{c'_E_new}) reduces to
\be\begin{split}\label{c'_E_simp}
c'_E = -2[(\mu_{R_1} - \mu_{L_1}+\mu_0)+(\mu_{R_2} - \mu_{L_2}+\mu_0)]-(9\mu_Q+\mu_{L_1}+\mu_{L_2}+\mu_{L_3}).
\end{split}\ee
In the above equation, the first, the second and the third parentheses correspond to the electron Yukawa reactions, the muon Yukawa reactions and the weak sphaleron processes, respectively. The parentheses vanish when their corresponding reactions are in chemical equilibrium. We let the parentheses evolve freely according to the evolution of their constituents when the evolution equations are solved numerically. 

\subsection{The Evolution Equation of the Hypermagnetic Field}\label{The Dynamical Equation of the Hypermagnetic Field}
The generalized diffusion equation of the hypermagnetic field and the generalized Ohm's law, derived from the AMHD equations are the following \cite{shiva},
\be \label{hypermagnetic}
\frac{\partial \textbf{B}_\textbf{Y}} {\partial t} = \frac{1}{\sigma}\nabla^2\textbf{B}_\textbf{Y} + \alpha_Y\nabla\times\textbf{B}_\textbf{Y},\ \ \ \ \ \ \ \ \ \ \ \ \ \ \ \ \ \ \ \ \ \ \ \ \ \ \ \ \ \ \ \ \ \ \ \ \ \ \ \ \ \ \ \ \ \ \ \  
\ee
\be\label{E_Y} 
\textbf{E}_\textbf{Y} = - \textbf{V}\times\textbf{B}_\textbf{Y} + \frac{\nabla\times\textbf{B}_\textbf{Y}}{\sigma} - \alpha_Y\textbf{B}_\textbf{Y}\ \ \ \textrm{where}\ \ \ \alpha_Y(T) = - c'_E\frac{g'^2}{8\pi^2\sigma}.
\ee
In the above equations, $\sigma\sim100T$ \cite{Arnold} is the hyperconductivity of the plasma, 
and $c'_E$ is given by Eq.\ (\ref{c'_E_simp}).
We choose the following simple nontrivial configuration for the hypermagnetic field
\be\label{wave}
Y_x=Y(t)\sin k_0z,\ \ \ \ \ Y_y=Y(t)\cos k_0z,\ \ \ \ \ Y_z =Y_0 =0,
\ee
which yields the hypermagnetic field amplitude $B_Y(t)=k_0Y(t)$. Substituting these into Eq.\ (\ref{hypermagnetic}), we 
obtain the evolution equation of $B_Y(t)$ in the form
\be \label{B_Y(t)}
\frac{dB_Y(t)}{dt} = B_Y(t)\left[-\frac{k_0^2}{\sigma(t)}-\frac{k_0 g'^2}{8\pi^2\sigma(t)}c'_E(t)\right].
\ee
It can be seen that the evolution of the hypermagnetic field is coupled to those of the chemical potentials (matter asymmetries) through $c'_E(t)$ as given by Eq.\ (\ref{c'_E_simp}). 
Let us now obtain the expression corresponding to the Abelian anomaly ($\sim \textbf{E}_\textbf{Y}.\textbf{B}_\textbf{Y}$) which appears in the evolution equations of the asymmetries in the next subsection. 
Using Eq.\ (\ref{E_Y}) and the simple configuration of the hypermagnetic field given by Eq.\ (\ref{wave}), we obtain
\be\label{E_Y.B_Y2} 
\textbf{E}_\textbf{Y}.\textbf{B}_\textbf{Y} = \frac{k_0}{\sigma}B_Y^2 - \alpha_Y B_Y^2.
\ee
Substituting the expression for $\alpha_Y$ given by Eq.\ (\ref{E_Y}) into the above equation, and using $\sigma = 100T$ 
leads to 
\be\label{E_Y.B_Y3}
\textbf{E}_\textbf{Y}.\textbf{B}_\textbf{Y} = \frac{B_Y^2}{100} \left[\frac{k_0}{T}+\frac{g'^2}{8\pi^2T}c'_E\right].
\ee
\subsection{The Evolution Equations of the Lepton and Baryon Asymmetries}\label{The Evolution Equations of the Lepton and Baryon Asymmetries}

As explained in Subsection \ref{Nonequilibrium and Almost Equilibrium Processes}, it is sufficient to solve the evolution equations only for the asymmetries of right-handed and left-handed electrons, and right-handed muons, to obtain all of the matter and Higgs asymmetries. We derive the dynamical equations for these three leptonic asymmetries taking into account the Abelian anomaly, the U$_\textrm{Y}$(1) Chern-Simons term, the chirality flip through Yukawa reactions, and the weak sphaleron processes. The violation of the lepton numbers via the Abelian anomaly is given by \cite{Long},
\bea\label{e_RL}
\partial_\mu j_{R_2}^\mu =\partial_\mu j_{R_1}^\mu &= -\frac{1}{4}(Y_R^2) \frac{g'^2}{16\pi^2}Y_{\mu\nu} {\tilde{Y}}^{\mu\nu} &= \frac{g'^2}{4\pi^2}(\textbf{E}_\textbf{Y}.\textbf{B}_\textbf{Y}),\\*\nonumber
 \partial_\mu j_{L_1}^\mu &= +\frac{1}{4}(Y_L^2) \frac{g'^2}{16\pi^2}Y_{\mu\nu} {\tilde{Y}}^{\mu\nu} &= - \frac{g'^2}{16\pi^2}(\textbf{E}_\textbf{Y}.\textbf{B}_\textbf{Y}),
\eea
where, $Y_{\mu\nu} {\tilde{Y}}^{\mu\nu}=-4\ \textbf{E}_\textbf{Y}.\textbf{B}_\textbf{Y}$ and the relevant hypercharges are 
\be 
Y_R = -2,\ \ \ Y_{L} = -1.
\ee
Using Eq.\ (\ref{asym-chem}), 
the system of dynamical equations for the leptonic asymmetries takes the form,\footnote{See Eq.\ (2.6) in Section 2.1 of Ref.\ \cite{Kamada} for the general form of the equations.}
\bea\label{lepton equations} 
\frac{d\eta_{R_1}}{dt} = +\frac{g'^2}{4\pi^2s}(\textbf{E}_\textbf{Y}.\textbf{B}_\textbf{Y}) - \Gamma_1(\eta_{R_1}-\eta_{L_1}+\frac{\eta_0}{2}),\ \ \ \ \ \ \ \ \ \ \ \ \ \ \ \\*\nonumber
\textrm{for}\ \ \ \ e_R\bar{\nu}_e \leftrightarrow \phi^{(-)}\ \ \ \textrm{and}\ \ \ e_R\bar{e}_L \leftrightarrow \tilde{\phi}^{(0)} ,\ \ \ \ \ \ \\*\nonumber
\frac{d\eta_{L_1}}{dt} = -\frac{g'^2}{16\pi^2s}(\textbf{E}_\textbf{Y}.\textbf{B}_\textbf{Y}) + \frac{\Gamma_1}{2}(\eta_{R_1}-\eta_{L_1}+\frac{\eta_0}{2})- \frac{\Gamma_{sph}}{2} \eta_E,\\*\nonumber
\textrm{for}\ \ \ \ \bar{e}_R{e}_L \leftrightarrow {\phi}^{(0)} ,\ \ \ \ \ \ \ \ \ \ \ \ \ \ \ \ \ \ \ \ \ \ \ \ \ \ \ \ \ \ \ \ \ \\*\nonumber
\frac{d\eta_{R_2}}{dt} = +\frac{g'^2}{4\pi^2s}(\textbf{E}_\textbf{Y}.\textbf{B}_\textbf{Y}) - \Gamma_2(\eta_{R_2}-\eta_{L_2}+\frac{\eta_0}{2}),\ \ \ \ \ \ \ \ \ \ \ \ \ \ \ \\*\nonumber
\textrm{for}\ \ \ \ \mu_R\bar{\nu}_{\mu} \leftrightarrow \phi^{(-)}\ \ \ \textrm{and}\ \ \ \mu_R\bar{\mu}_L \leftrightarrow \tilde{\phi}^{(0)} ,\ \ \ \ \  
\eea
In the above equations, the terms containing $\textbf{E}_\textbf{Y}.\textbf{B}_\textbf{Y}$ originate from the Abelian anomaly equations (\ref{e_RL}). The rate associated with the direct and inverse Higgs decay processes via Yukawa interactions for the \textit{i}th-generation leptons, as estimated in Eq.\ (2.26) of Ref.\ \cite{Kamada} is
$\Gamma_i \sim 10^{-2}h_i^2T/8\pi = \Gamma_i^0/t_{EW}\sqrt{x}$, where $h_i$ is the relevant Yukawa coupling constant. For $h_1=2.8\times10^{-6}$ and $h_2=5.8\times10^{-4}$ as given in Appendix B of Ref.\ \cite{Fujita}, we obtain $\Gamma_1^0 \simeq 11.1$, and $\Gamma_2^0 \simeq 4.8\times10^5$. Moreover, the variable $x = t/t_{EW} = (T_{EW}/T )^2$ due to the Friedmann law, $t_{EW} = M_0/2T_{EW}^2$ and $M_0 = M_{\textrm{Pl}}/1.66\sqrt{g^*}$. 
The factor 1/2 multiplying the rates $\Gamma_i$ and $\Gamma_\textrm{sph}$ in the second line is due to the equivalent rates of reaction branches for left-handed electron and neutrino. 
The effect of the weak sphaleron processes on the evolution is investigated by substituting their corresponding term $(\Gamma_{\textrm{sph}}/{2}) \eta_E$ in the dynamical equation of left-handed electron asymmetry \cite{Khlebnikov,Kamada}. In this term, $\eta_E=\frac{T^2}{6s} c_E=9\eta_Q+\eta_{L_1}+\eta_{L_2}+\eta_{L_3}$, and the rate of the weak sphalerons is $\Gamma_\textrm{sph} \simeq 25\alpha_w^5T=\Gamma_\textrm{sph}^0/t_{EW}\sqrt{x}$, where $\alpha_W\simeq 3.17 \times 10^{-2}$ and $\Gamma_\textrm{sph}^0=2.85\times10^9$\cite{Long,Dvornikov}.

Defining $y \equiv 10^4\mu/T$, the fermion and the boson asymmetries can be written as $\eta = 10^{-4} y T^3c/6s$. Then, using Eq.\ (\ref{E_Y.B_Y3}), Eqs.\ (\ref{lepton equations}) can be rewritten in terms of the dimensionless chemical potentials $y$ in the form, 
\be\begin{split}\label{y_equations}
\frac{dy_{R_1}}{dx} = \left[B_0 x^{\frac{1}{2}}-A_0y_T\right]\left(\frac{B_Y(x)}{10^{20}\mbox{G}}\right)^2x^{\frac{3}{2}} -\frac{\Gamma_1^0}{\sqrt{x}}(y_{R_1}-y_{L_1}+y_0),\ \ \ \ \ \ \ \ \ \ \ \ \ \ \ \ \ \ \ \ \cr
\frac{dy_{L_1}}{dx} = \frac{-1}{4}\left[B_0 x^{\frac{1}{2}}-A_0y_T\right]\left(\frac{B_Y(x)}{10^{20}\mbox{G}}\right)^2x^{\frac{3}{2}}+\frac{\Gamma_1^0}{2\sqrt{x}}(y_{R_1}-y_{L_1}+y_0)-\frac{\Gamma_\textrm{sph}^0}{2\sqrt{x}}y_E,\cr
\frac{dy_{R_2}}{dx} = \left[B_0 x^{\frac{1}{2}}-A_0y_T\right]\left(\frac{B_Y(x)}{10^{20}\mbox{G}}\right)^2x^{\frac{3}{2}} -\frac{\Gamma_2^0}{\sqrt{x}}(y_{R_2}-y_{L_2}+y_0),\ \ \ \ \ \ \ \ \ \ \ \ \ \ \ \ \ \ \ \ \cr
\end{split}\ee
where,
\be\begin{split}\label{yE-yT}
y_E=9y_Q+y_{L_1}+y_{L_2}+y_{L_3},\ \ \ \ \ \ \ \ \ \ \ \ \ \ \ \ \ \ \ \ \ \ \ \ \ \ \ \ \ \ \cr
y_T = (y_{R_1} - y_{L_1}+y_0)+(y_{R_2} - y_{L_2}+y_0)+\frac{1}{2}y_E.\ \ \ \     
\end{split}\ee
In Eqs.\ (\ref{y_equations}), $B_0 = 25.6\left(\frac{k_0}{10^{-7}T_{EW}}\right)$ and $A_0 = 77.6$ are chosen to normalize the hypermagnetic field amplitude at $10^{20}$G.
Let us now rewrite Eqs.\ (\ref{other asymmetries}) in terms of the dimensionless chemical potentials $y$, 
\be\begin{split}\label{y0-yL2}
y_Q=\frac{y_{R_1}+2y_{L_1}}{4},\ \ \ \ \ \ \ \ \ \ \ \ \ \ \ \ \ \ y_{L_2}=\frac{y_{R_1}+2y_{L_1}-y_{R_2}}{2},\ \ \cr
y_{L_3}=\frac{46y_{L_1}+24y_{R_1}+y_{R_2}}{68},\ \ \ \ \ \ y_0=\frac{4y_{R_1}+2y_{L_1}+3y_{R_2}}{68}.
\end{split}\ee
Using Eq.\ (\ref{mu_B}) (or equivalently $y_B=12y_Q$) and the expression given for $y_Q$ in Eqs.\ (\ref{y0-yL2}), we can also obtain the baryonic dimensionless chemical potential as 
\be\label{y_B} 
y_B=3(y_{R_1}+2y_{L_1}). 
\ee
Substituting $y_Q$, $y_{L_2}$ and $y_{L_3}$ from Eqs.\ (\ref{y0-yL2}) into the expression for $y_E$ as given by Eq.\ (\ref{yE-yT}), we obtain
\be\label{yE}
y_E=\frac{211y_{R_1}+488y_{L_1}-33y_{R_2}}{68}.\ \ 
\ee 
Using Eqs.\ (\ref{y0-yL2}) and (\ref{yE}), we simplify the expression for $y_T$ as given by Eq.\ (\ref{yE-yT}) and obtain
\be\begin{split}\label{yT}
y_T = \frac{295y_{R_1}+224y_{L_1}+183y_{R_2}}{136}.
\end{split}\ee
Rewriting Eq.\ (\ref{B_Y(t)}) in terms of $x$ and $y_T$ as given above leads to
\be \label{B_Y(x)}
\frac{dB_Y}{dx} = 3.5\left(\frac{k_0}{10^{-7}T_{EW}}\right)\left[\frac{y_T}{\pi}-0.1\left(\frac{k_0}{10^{-7}T_{EW}}\right)\sqrt{x})\right]B_Y(x).
\ee
We substitute $y_0$ and $y_{L_2}$ given by Eqs.\ (\ref{y0-yL2}), and $y_E$ and $y_T$ given by Eqs.\ (\ref{yE}) and (\ref{yT}), into the set of dynamical equations as given by Eqs.\ (\ref{y_equations}) and (\ref{B_Y(x)}) with the equilibrium initial conditions as described in Section \ref{Equilibrium Conditions} to obtain a minimal set of self-consistent evolution equations for the matter asymmetries and the hypermagnetic field.


\begin{figure} 
  \includegraphics[width=65mm]{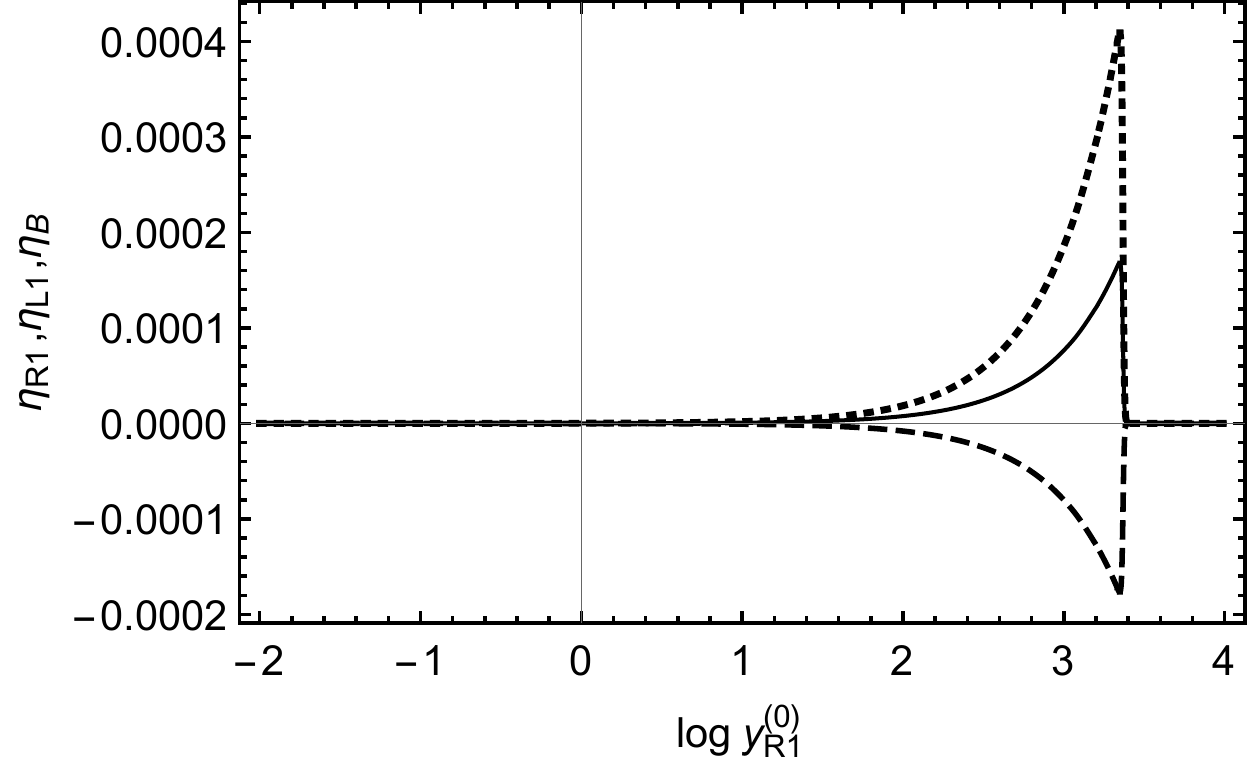}
  \hspace{2mm}
  \includegraphics[width=65mm]{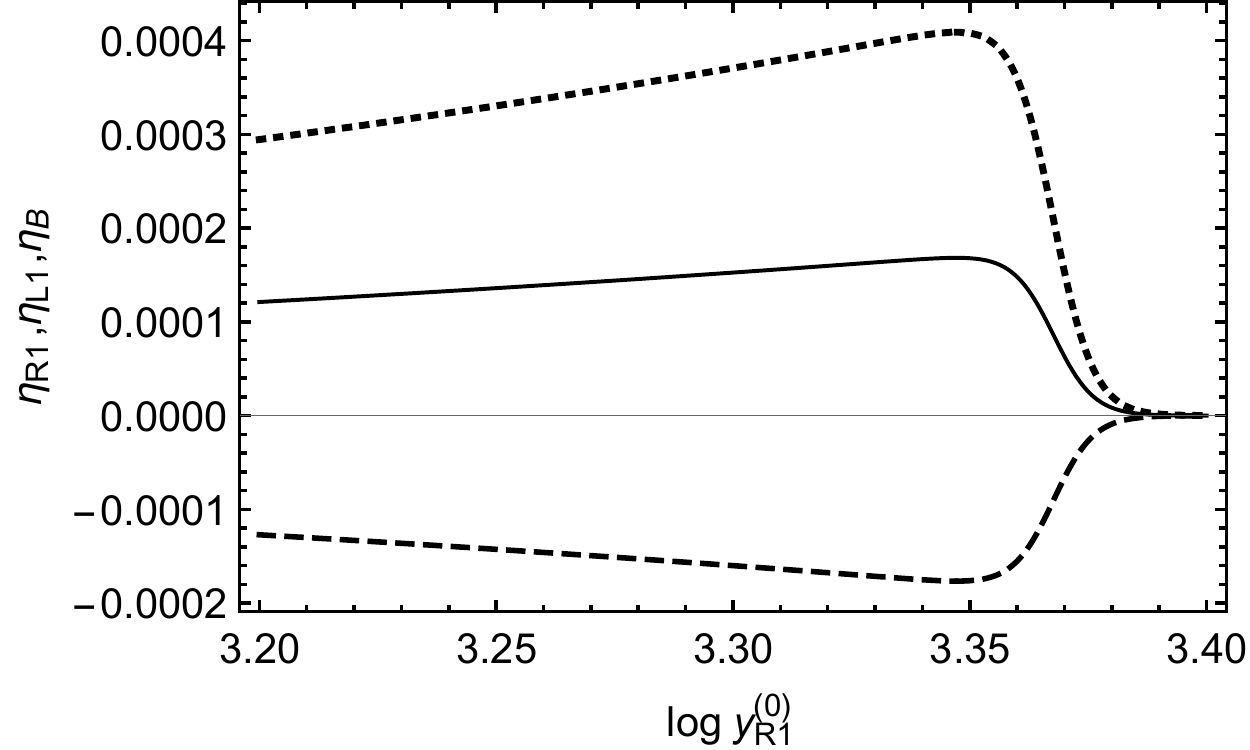}
  \includegraphics[width=65mm]{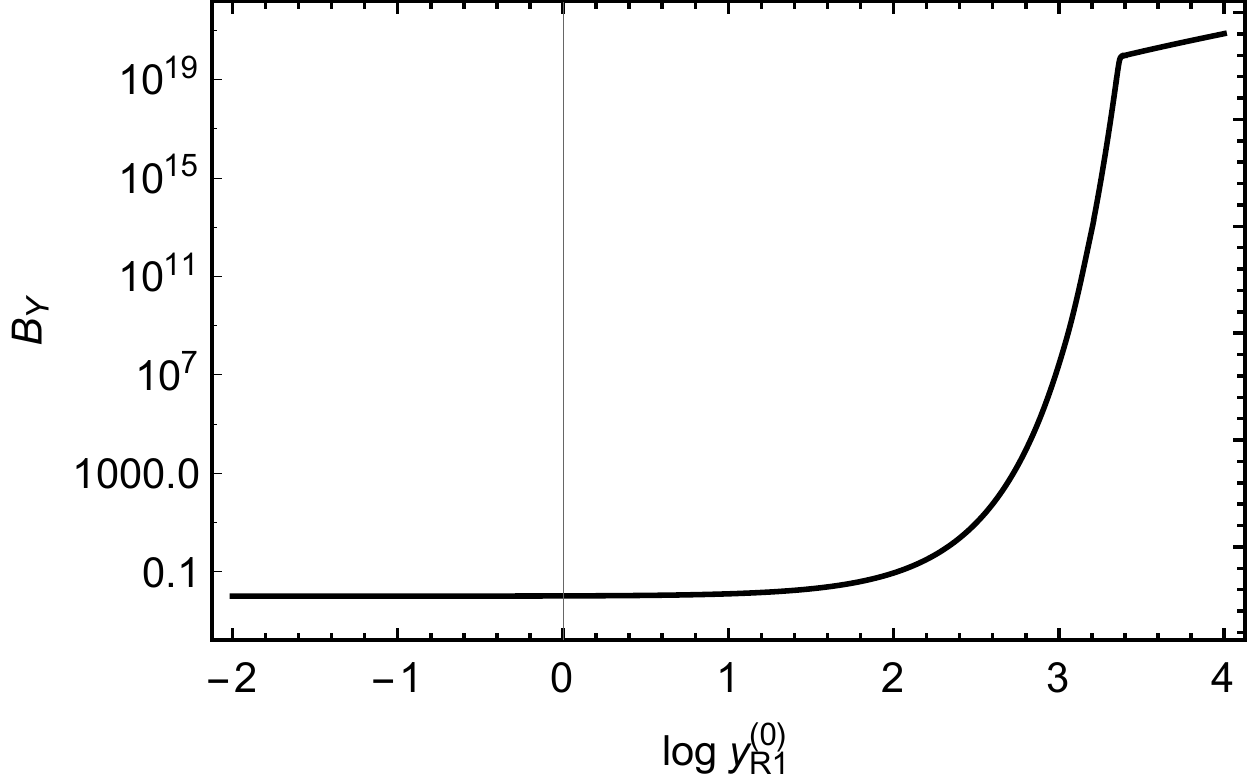}
  \hspace{2mm}
  \includegraphics[width=65mm]{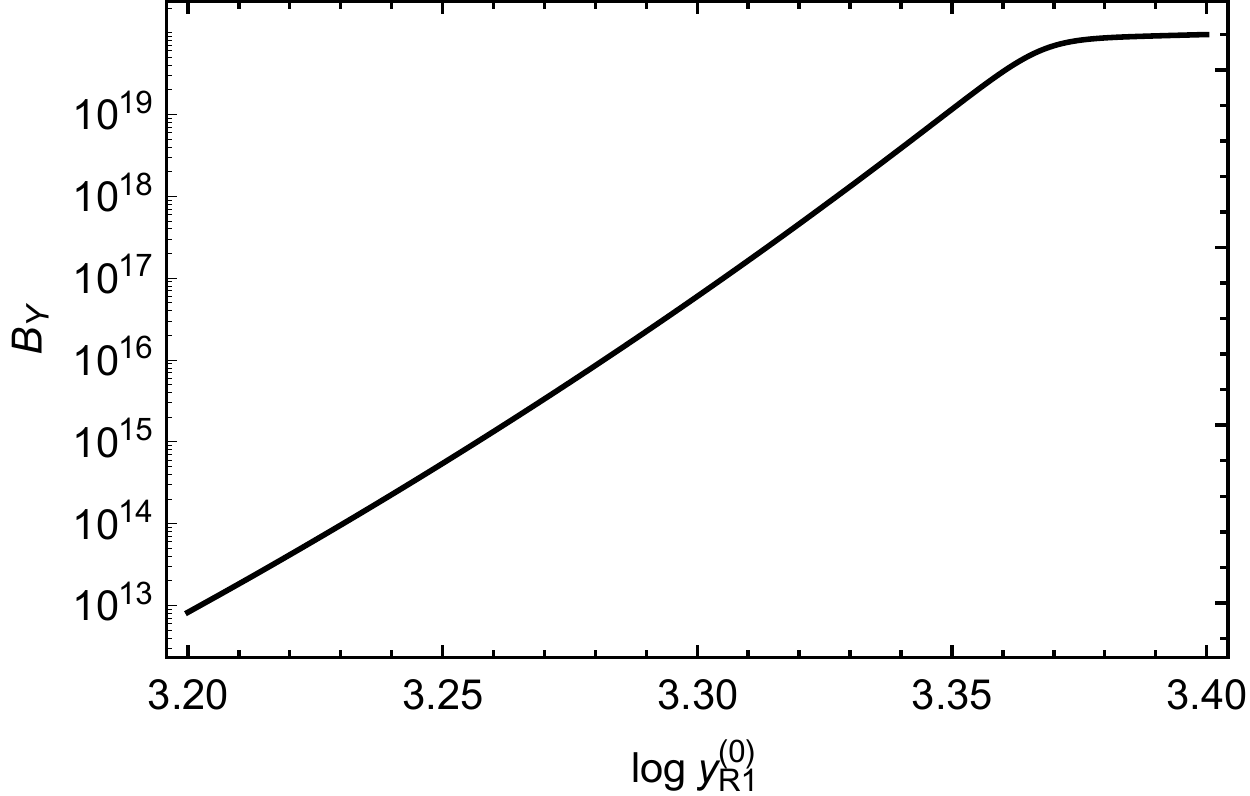}
\caption{The final values of our dynamical variables (at $T=T_{EW}$) for (Top): the asymmetries of right-handed electrons $\eta_{R_1}$ (dotted line), baryons $\eta_B$ (solid line), left-handed electrons $\eta_{L_1}$ (dashed line), and 
(Bottom): hypermagnetic field amplitude $B_Y$, for $B_Y^{(0)}=10^{-2}$G, $k=0.02$ ($k_0=2\times 10^{-9}T_{EW}$), $c_{\Gamma_1}=0.02$, and $\log(y_{R_1}^{(0)})$ varies from $-2$ to $4$. The maximum relative error for these graphs is of the order of $10^{-12}$.
} \label{four-}
\end{figure}

\begin{figure} 
  \includegraphics[width=65mm]{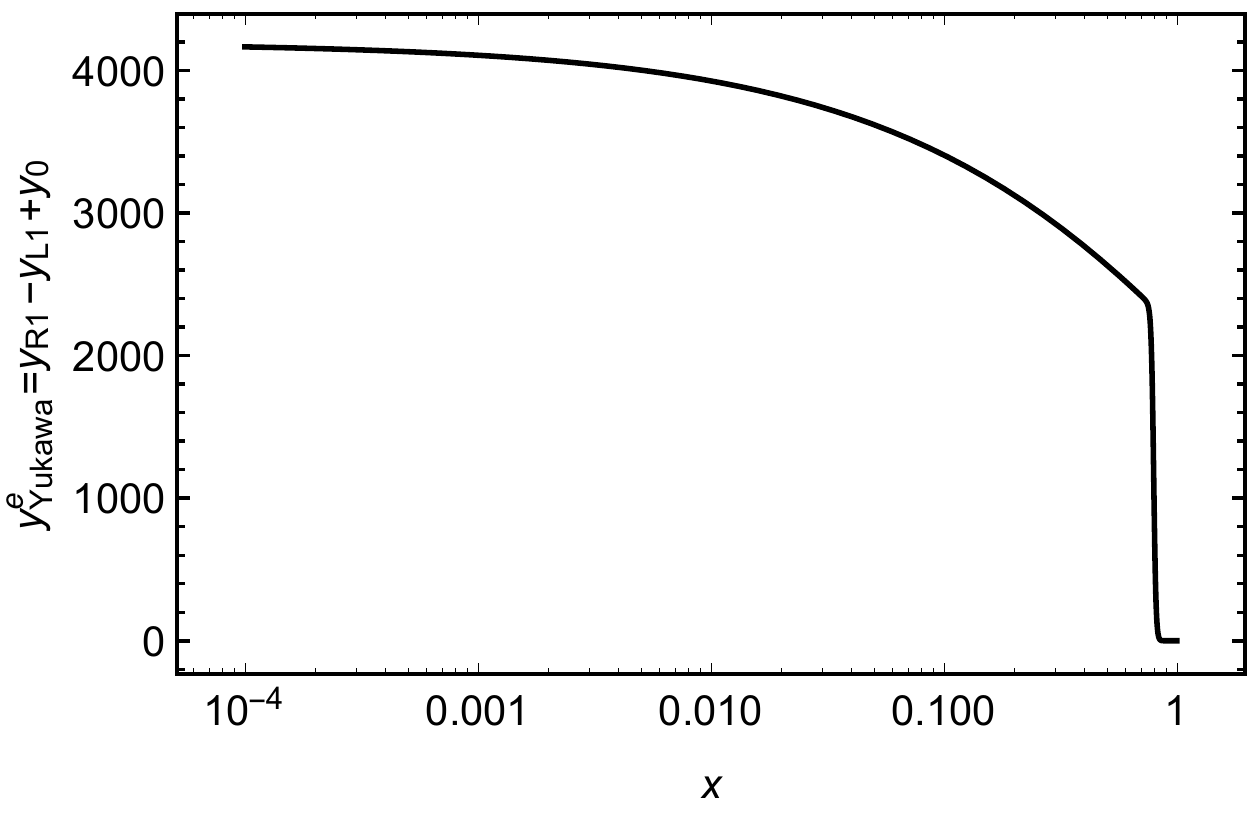}
  \hspace{2mm}
  \includegraphics[width=65mm]{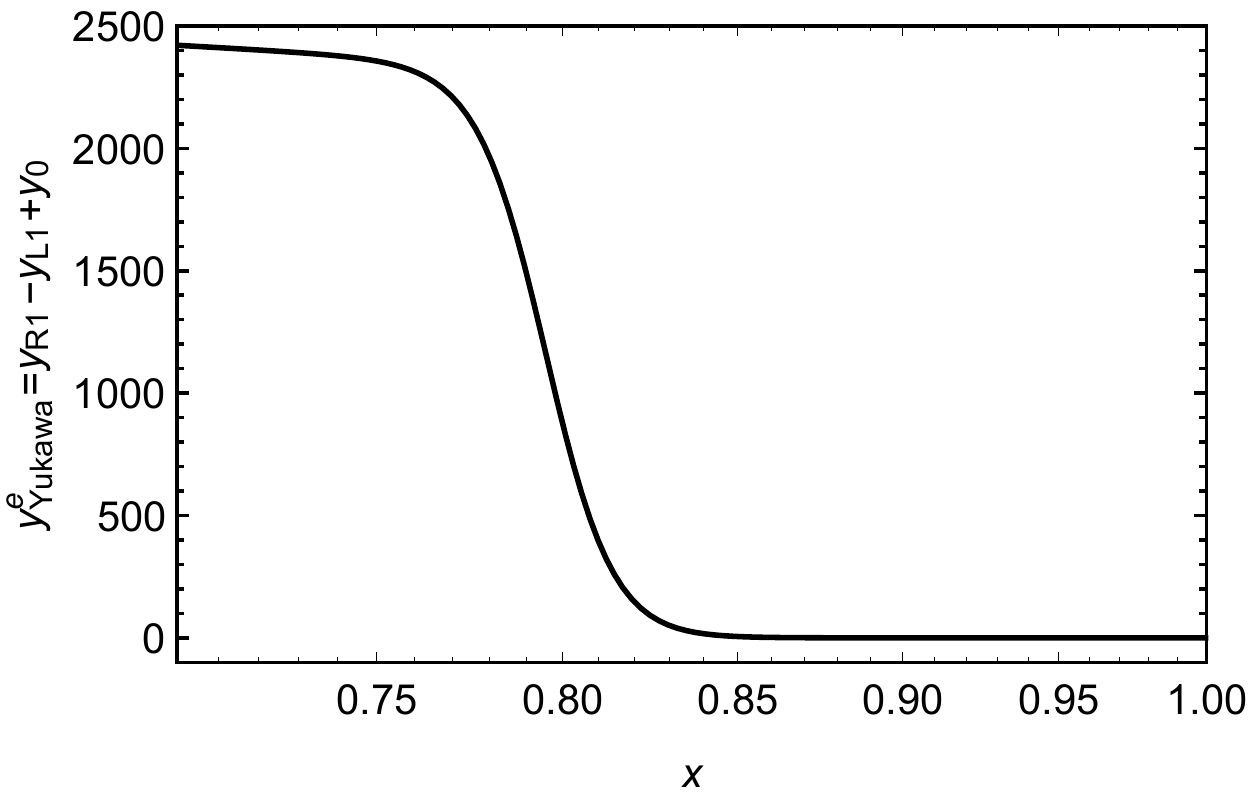}
  \includegraphics[width=65mm]{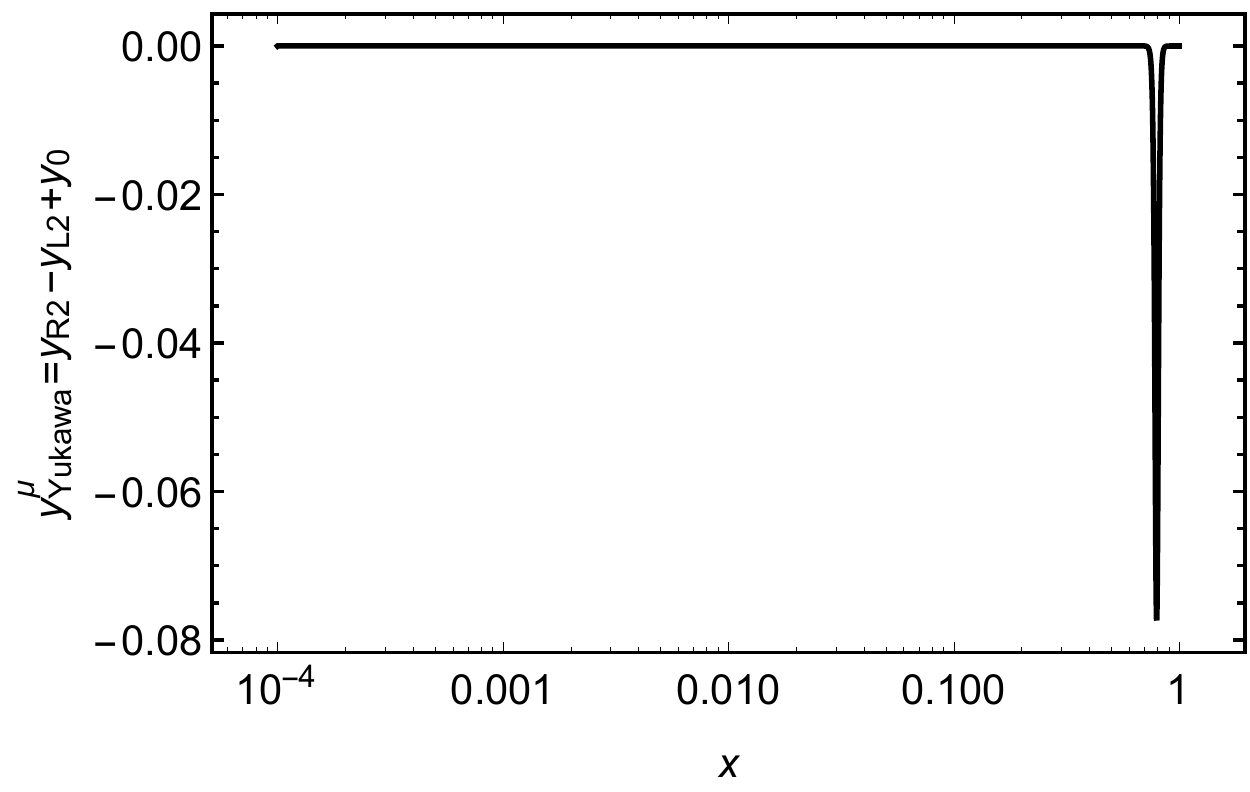}
  \hspace{2mm}
  \includegraphics[width=65mm]{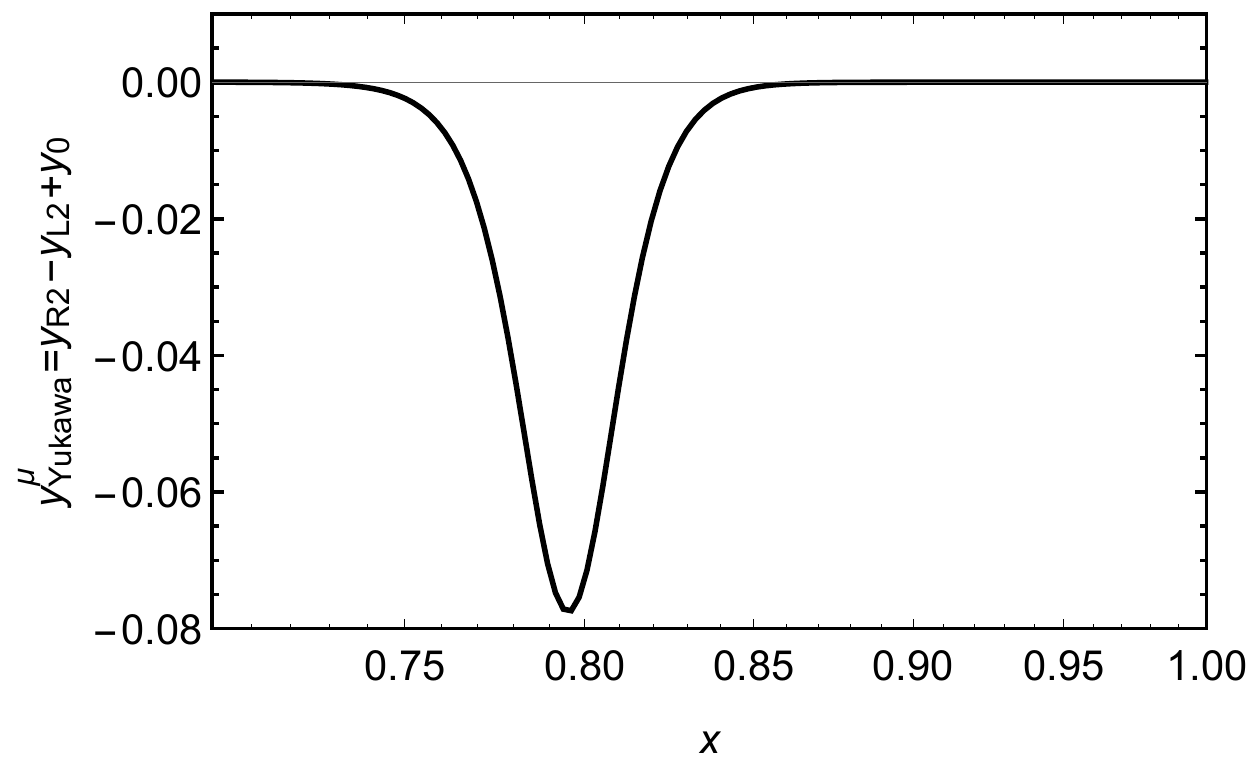}
  \includegraphics[width=65mm]{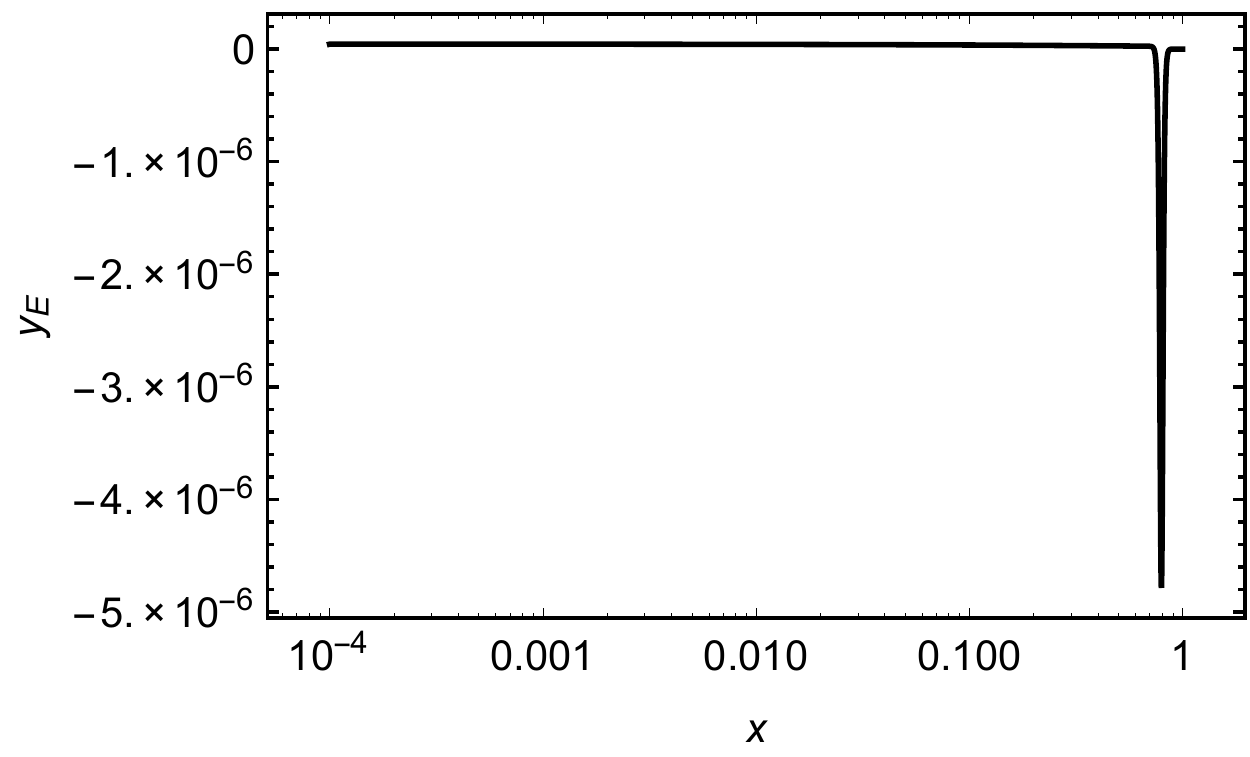}
  \hspace{2mm}
  \includegraphics[width=65mm]{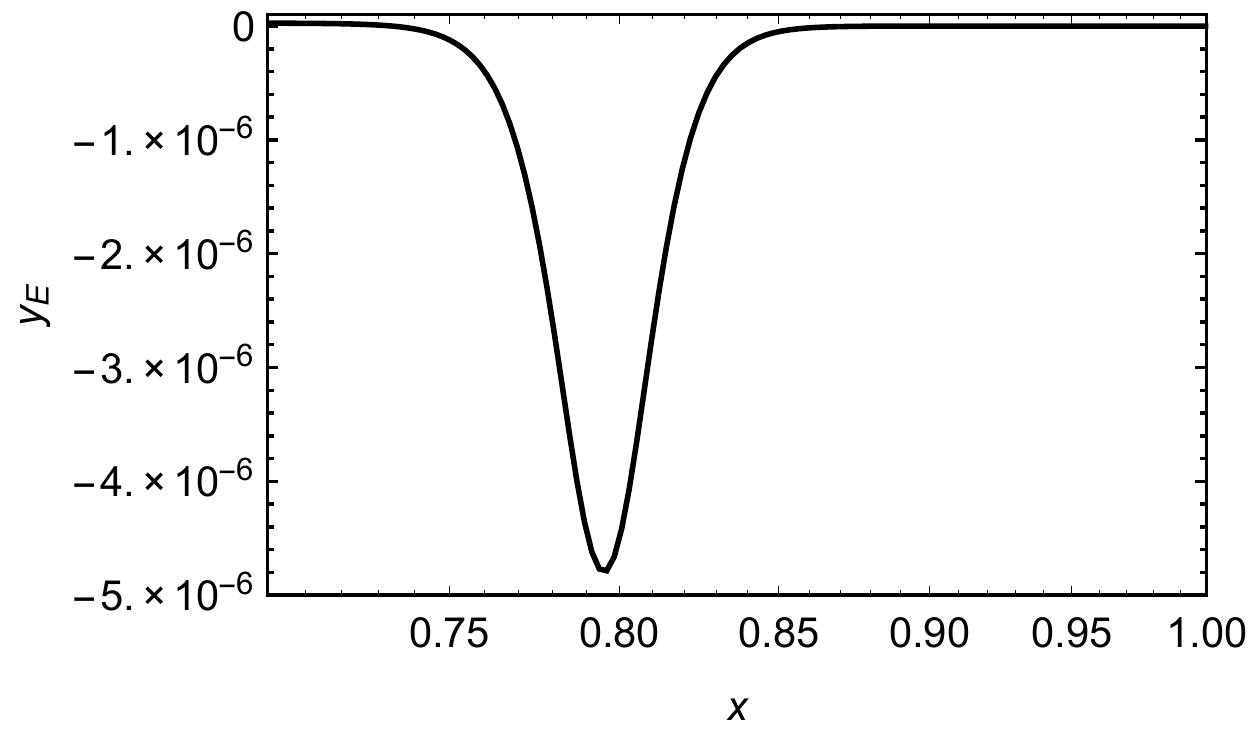}
\caption{The time plots of (top): $y_{\mbox{\tiny{Yukawa}}}^e=y_{R_1}-y_{L_1}+y_0$, (middle): $y_{\mbox{\tiny{Yukawa}}}^{\mu}=y_{R_2}-y_{L_2}+y_0$, and (bottom): $y_E$, representing the amounts of falling out of chemical equilibrium for the electron Yukawa reactions, the muon Yukawa reactions and the weak sphaleron processes, respectively, for $B_Y^{(0)}=10^{-2}$G, $k=0.02$ ($k_0=2\times 10^{-9}T_{EW}$), $c_{\Gamma_1}=0.02$, and $\log(y_{R_1}^{(0)})=3.45$. The maximum relative error for these graphs is of the order of $10^{-14}$.\\
}\label{3.45ys}
\end{figure}

\begin{figure}
  \includegraphics[width=65mm]{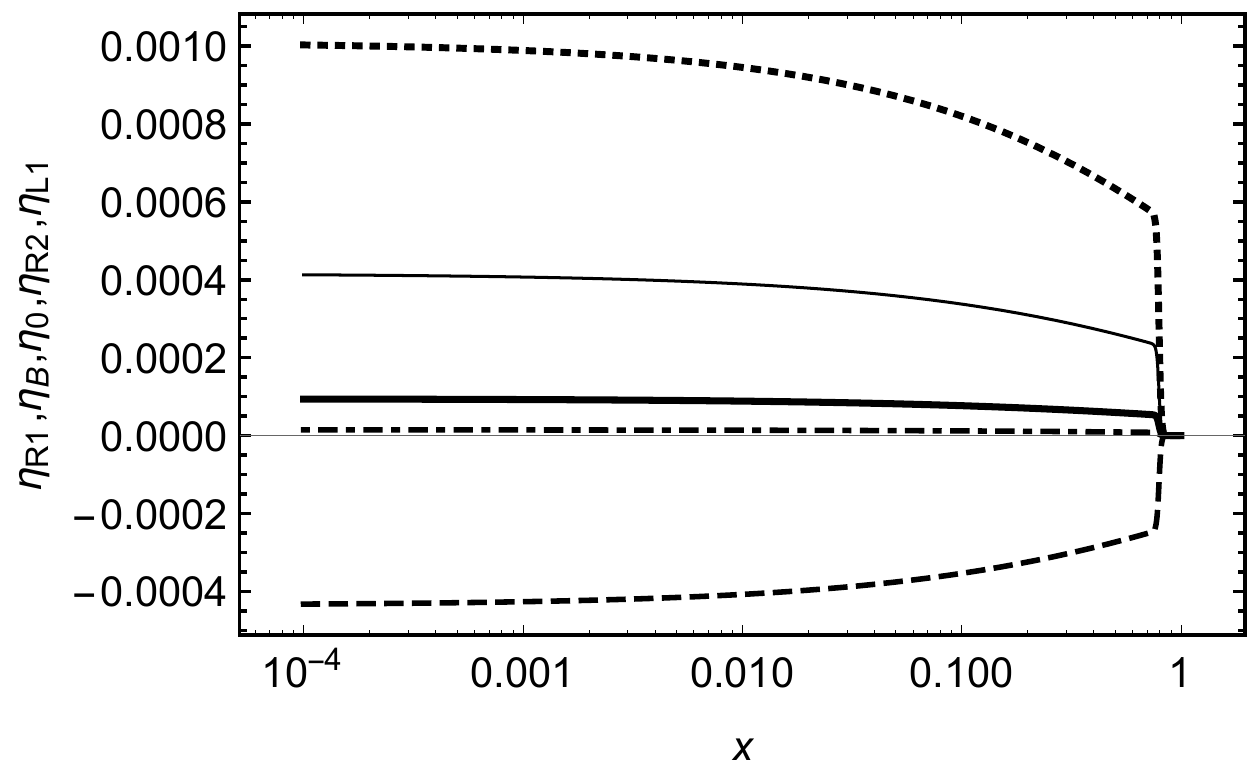}
  \hspace{2mm}
  \includegraphics[width=65mm]{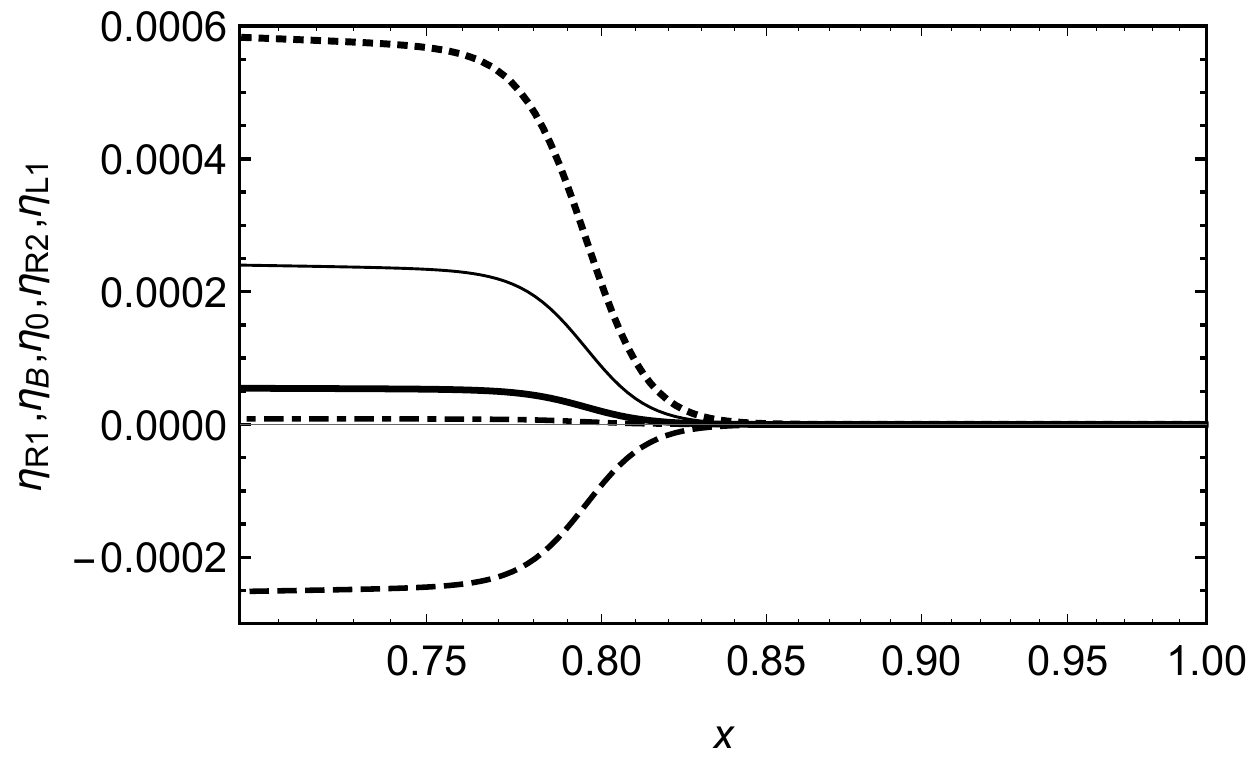}
  \includegraphics[width=65mm]{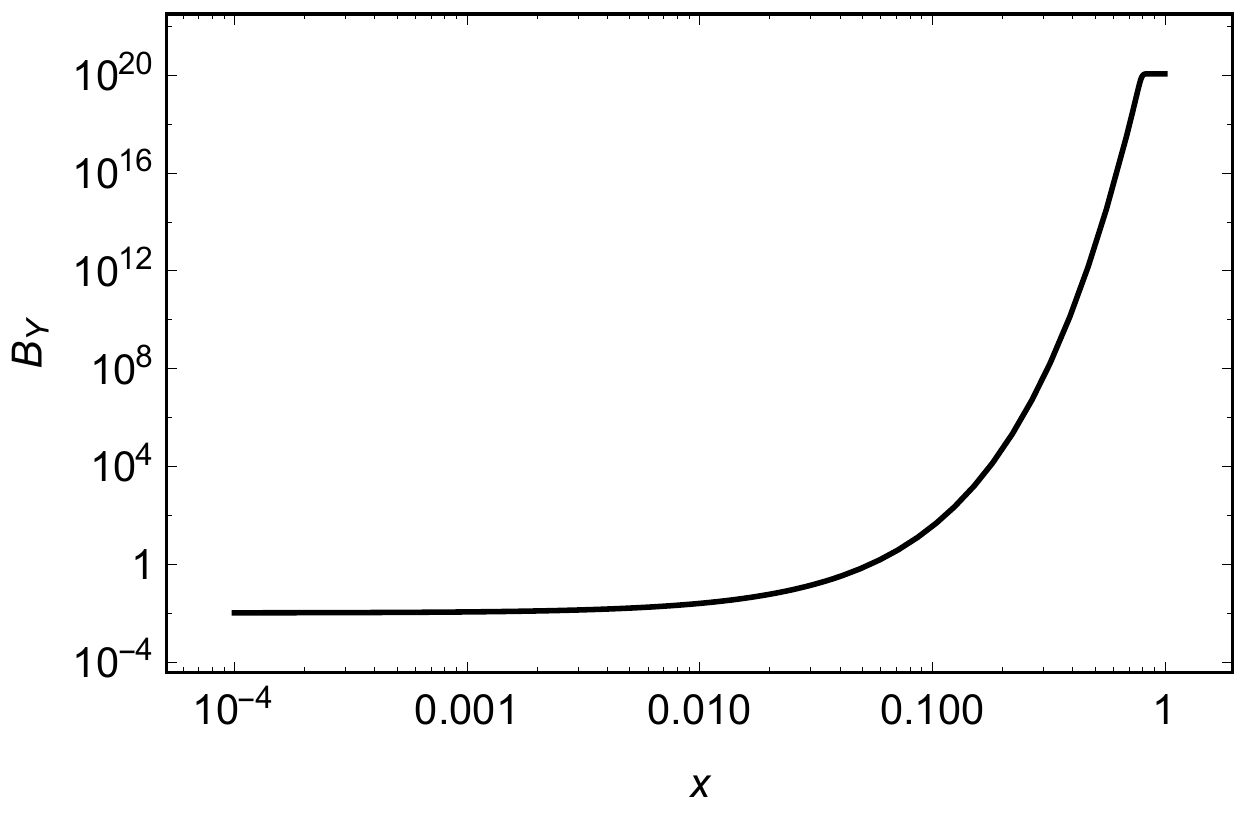}
  \hspace{2mm}
  \includegraphics[width=65mm]{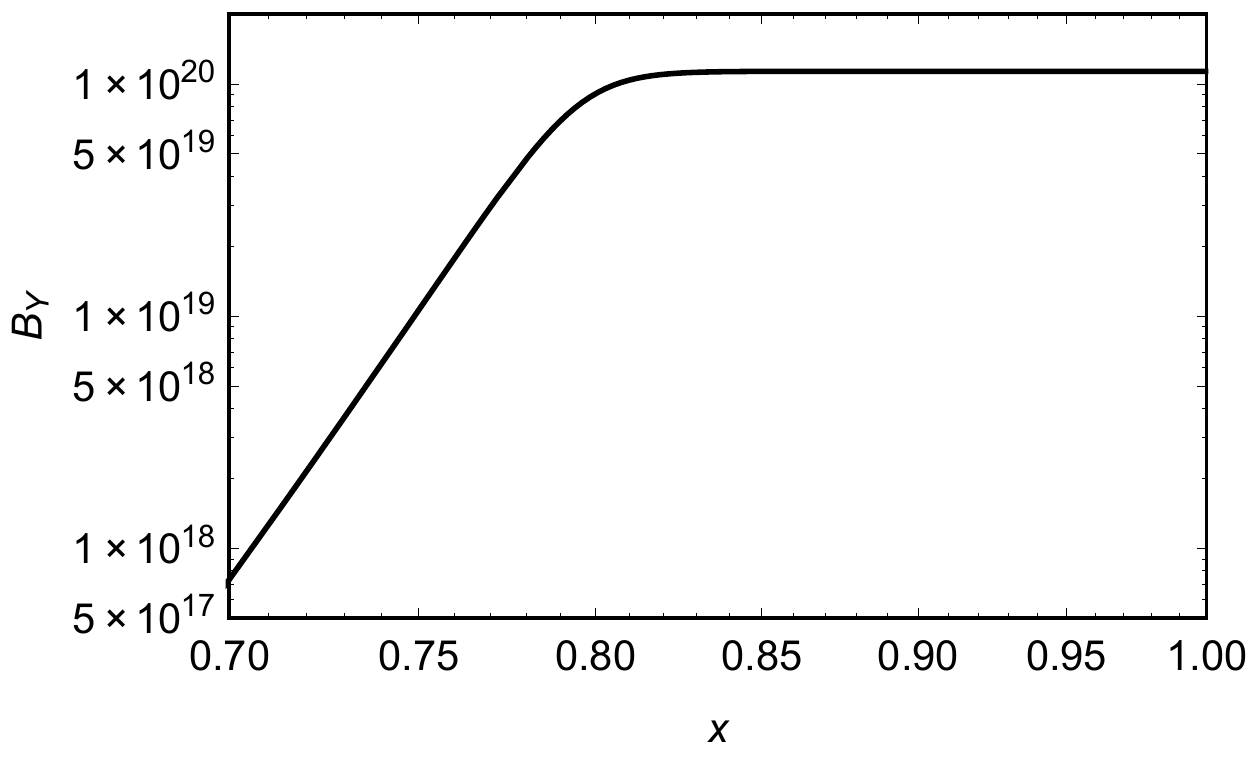}
\caption{The time plots for (top): the asymmetries of right-handed electrons $\eta_{R_1}$ (dotted line), baryons $\eta_B$ (thin line), Higgs bosons $\eta_0$ (thick line), right-handed muons $\eta_{R_2}$ (dotdashed line), left-handed electrons $\eta_{L_1}$ (dashed line), and 
(bottom): hypermagnetic field amplitude $B_Y$, for $B_Y^{(0)}=10^{-2}$G, $k=0.02$ ($k_0=2\times 10^{-9}T_{EW}$), $c_{\Gamma_1}=0.02$, and $\log(y_{R_1}^{(0)})=3.45$. The maximum relative error for these graphs is of the order of $10^{-14}$.\\
}\label{3.45etas}
\end{figure}

\section{Results}\label{Results}
In Section 2, we discussed the equilibrium conditions in the electroweak plasma. There, we obtained a set of consistency relationships between the chemical potentials including six relations between $\mu_Q, \mu_{R_1}, \mu_{L_1}, \mu_{L_2}, \mu_{L_3}$ and $\mu_0$ (see Eqs.\ (\ref{sphaleron}),(\ref{charge}) and (\ref{charge2})). As stated in the paragraph below Eqs.\ (\ref{charge2}), in this study we choose $c_1,c_2$ and $c_3$ appearing in Eqs.\ (\ref{charge2}) to be zero. When we want to solve the evolution equations starting with non-zero matter asymmetries, it is then sufficient to choose a non-zero value for the right-handed electron asymmetry, $\eta_{R1}=c_{R1} $. We then solve the consistent set of equations to get the initial values for all other chemical potentials in terms of $c_{R_1}$. To do this, we first solve the six equations to obtain the initial values for the aforementioned six chemical potentials, then obtain the initial values of other chemical potentials through their relationship with the known ones (see Eqs.\ (\ref{Yukawa1}),(\ref{Yukawa}) and (\ref{mu_B})). Since, the above arguments remain true for the normalized chemical potentials ($y$) as well, we use variable $y$ instead of $\mu$ in what follows and replace $c_{R_1}$ by $y_{R_1}^{(0)}$ which is the initial value for the normalized chemical potential of the right-handed electron. Therefore, in the following subsections, choosing $y_{R_1}^{(0)}=0$ leads to the condition of zero initial normalized chemical potentials (or equivalently matter asymmetries), while choosing nonzero $y_{R_1}^{(0)}$ results in nonzero initial values for the normalized chemical potentials (or equivalently matter asymmetries) in terms of $y_{R_1}^{(0)}$.


In this section we solve our four coupled evolution equations, with initial conditions respecting the conditions stated above. We divide the initial conditions into two main categories. First we assume the presence of large initial matter asymmetries and a tiny seed of hypermagnetic field and explore this problem with particular emphasis on the possibility of generating large hypermagnetic fields. Then we do the opposite, that is we start with a large hypermagnetic field and assume zero initial matter asymmetries and explore this problem with particular emphasis on the possibility of generating large matter asymmetries.

\begin{table}[ht]
\caption{$B_Y^{(0)}=10^{-2}$G, $k=0.02$ and $c_{\Gamma_1}$=0.02}
\label{table:c0.02}
\centering
\begin{tabular}{c c c c c}
\hline 
& $\log(y_{R_1}^{(0)})$ & $\eta_B(1)$ & $B_Y(1)$ \\[0.5ex]
\hline
& 2.85 & $5.42\times10^{-5}$ & $4.58\times10^{4}$G \\
& 2.95 & $6.82\times10^{-5}$ & $2.43\times10^{6}$G \\
& 3.05 & $8.58\times10^{-5}$ & $3.61\times10^{8}$G \\
& 3.15 & $1.08\times10^{-4}$ & $1.95\times10^{11}$G \\
& 3.25 & $1.36\times10^{-4}$ & $5.40\times10^{14}$G \\
& 3.35 & $1.68\times10^{-4}$ & $1.15\times10^{19}$G \\
& 3.45 & $6.50\times10^{-10}$ & $1.13\times10^{20}$G \\
& 3.55 & $6.52\times10^{-10}$ & $1.61\times10^{20}$G \\
& 3.65 & $6.53\times10^{-10}$ & $2.28\times10^{20}$G \\
& 3.75 & $6.54\times10^{-10}$ & $3.19\times10^{20}$G \\
& 3.85 & $6.54\times10^{-10}$ & $4.43\times10^{20}$G \\[1ex]
\hline
\end{tabular}
\end{table}



\subsection{Hypermagnetic field growth via matter asymmetries}\label{Hypermagnetic field growth via matter asymmetries}
In this subsection, we assume $c_1=c_2=c_3=0$ but choose nonzero initial values for right-handed electron asymmetry $y_{R_1}^{(0)}$ leading to nonzero initial matter asymmetries which can be obtained in terms of $y_{R_1}^{(0)}$ as explained earlier. Here, we want to investigate whether these matter asymmetries are able to grow a tiny seed of the hypermagnetic field in the presence of the weak sphalerons. Since the hypermagnetic field is not strong, it cannot make the Yukawa processes fall out of equilibrium to prevent the washout of the asymmetries by the weak sphalerons, as proposed in the Introduction. Therefore, the weak sphalerons wash out the asymmetries very quickly and no asymmetry remains to strengthen the hypermagnetic field at later times, specially in the time region $10^{-1}<x<1$ close to the EWPT. In fact, below $T_{RL}$, the electron chirality flip rate is high and the right-handed electrons transform into left-handed ones quickly, then the weak sphalerons transform them into antiquarks to be annihilated by the quarks. If the rate of the electron chirality flip processes was smaller, the rate of washout would be lower in the critical time region $10^{-4}<x<10^{-1}$, and therefore part of the asymmetries could be saved till near the EWPT to grow the hypermagnetic field.  

The first calculation of the electron chirality flip rate was performed by the authors of the first paper of Ref.\ \cite{Campbell} (CDEO), where the $2\to 1$ inverse Higgs decay was taken into account. Then, the authors of the third paper of Ref.\ \cite{Campbell} numerically calculated the rate and compared it with the approximate calculation of CDEO for some values of $m_H/T$ (see Table 2 of this reference). Recently, the authors of Ref.\ \cite{Bodeker} numerically calculated the rate of $1\leftrightarrow 2$ (inverse) Higgs decays, by taking into account the thermal fermion masses as well as the final state distribution function, which were neglected by CDEO (see dot-dashed green curve in Figure 5 of Ref.\ \cite{Bodeker}). It can be seen that the rate is about $(2-6) \times10^{-5} h_e^2T$, in the temperature range $1$TeV$<T<10$TeV. There are a number of assumptions implicit in these approximations of the rate, and it is not clear that they remain valid especially around the EWPT. The rate has a complicated dependence on the left-handed and right-handed electron masses as well as the Higgs mass. The growth of the Higgs condensate during the electroweak crossover can affect these masses (see Appendix B of Ref.\ \cite{Kamada}). Therefore, there is ambiguity in the value of this rate, especially in the time region $10^{-1}<x<1$.\footnote{The electron chirality flip rate $\Gamma_0(1-x)/2\sqrt{x}$ or $2 \times10^{-3} h_e^2T(1-(\frac{T_{EW}}{T})^2)$ approximately calculated by CDEO and used in Refs.\ \cite{shiva,Dvornikov}, has much smaller value than $\Gamma_1^0/2\sqrt{x}$ or $10^{-2} h_e^2T/8\pi$ as estimated in Ref.\ \cite{Kamada}, in the time region $10^{-1}<x<1$ especially when $x\to 1$. However, for the critical time region $10^{-4}<x<10^{-1}$, the latter is much more favorable, since the parameter $\Gamma_1^0=11$ is one order of magnitude smaller than $\Gamma_0=121$. Furthermore, the order of magnitude of the latter rate is closer to that of the recently calculated rate of Ref.\ \cite{Bodeker} in this time region.} In this work, we use the rates as estimated in Ref.\ \cite{Kamada}. Since, the rate of the electron Yukawa processes is an important parameter with a key role and there is an uncertainty in it, we multiply the rate by a varying adjustable parameter $c_{\Gamma_1}$ in order to investigate the dependence of the aforementioned scenario to the electron chirality flip rate, especially in the critical time region $10^{-4}<x<10^{-1}$.

In this subsection, we explore whether it is possible to grow a very weak seed of the hypermagnetic field with an initial amplitude of $B_Y^{(0)}=10^{-2}$G, i.e. at $T=T_{RL}$ or $x_i=\frac{t_{RL}}{t_{EW}}=(\frac{T_{EW}}{T_{RL}})^2=10^{-4}$, to a final amplitude, i.e. at $T=T_{EW}$ or $x_f=1$, as large as $B_Y(1)\sim 10^{20}$G by assuming the presence of large initial matter asymmetries, and at the same time, obtain the final baryonic asymmetry of $\eta_B(1)\simeq 6\times 10^{-10}$, namely the BAU (Baryonic Asymmetry of the Universe), at $T=T_{EW}$.\footnote{For simplicity, we are assuming these values do not change substantially at lower temperatures. Usually BAU is referred to as its value at $T=T_{\textrm{Hadronization}}\approx200\mbox{MeV}$. To get to this temperature, the Universe has to go through the electroweak phase transition, as well (see Ref. \cite{Abedi}).}

We explore the sensitivity of our results to the electron chirality flip rate ${\Gamma_1}$, by multiplying it with an adjustable parameter $c_{\Gamma_1}$.\footnote{Although we use the roughly estimated rates of Ref.\ \cite{Kamada} for the aforementioned processes, our investigations show that the main results 
are insensitive to these rates. 
Nevertheless, the electron Yukawa reaction, singled out as described above, plays a more dominant role due to its very small Yukawa coupling. We have also found that the mentioned insensitivity is due to the presence of the $\textrm{U}_{\textrm{Y}}(1)$ Chern-Simons term and in its absence, the results become sensitive to the rates.} We also define $k=\frac{k_0}{10^{-7}T_{EW}}$ as the normalized wave number of the hypermagnetic field. First, for $B_Y^{(0)}=10^{-2}$G, $k=0.02$, and $c_{\Gamma_1}=0.02$, we change the initial matter asymmetries by changing the value of $y_{R_1}^{(0)}$ as explained in Section \ref{Equilibrium Conditions}, then solve the evolution equations and obtain the final values of baryonic asymmetry $\eta_B(1)$ and the hypermagnetic field amplitude $B_Y(1)$, and present the results in Table \ref{table:c0.02} and Figure \ref{four-}. It can be seen that, for $\log(y_{R_1}^{(0)}) \lesssim 3.35$,
$\eta_B(1)$ increases and $B_Y(1)$ grows almost exponentially.
However, for $\log(y_{R_1}^{(0)})$ exceeding $3.35$, $\eta_B(1)$ decreases severely then saturates; while, $B_Y(1)$ increases with a very much smaller rate. 
More importantly, just after the sudden change, i.e. at  $\log(y_{R_1}^{(0)})=3.45$, $\eta_B(1) \simeq 6.5\times10^{-10}$ and $B_Y(1)\simeq 10^{20}$G are obtained.

The time plots corresponding to $\log(y_{R_1}^{(0)})=3.45$
are shown in Figures \ref{3.45ys} and \ref{3.45etas}. 
The plots show a great instability for $0.75 \lesssim x \lesssim 0.85$.
In this time interval, the asymmetries decrease but the hypermagnetic field amplitude increases, then both saturate as shown in Figure \ref{3.45etas}. Moreover, Figure \ref{3.45ys} shows that the amount of falling out of chemical equilibrium is very small for the muon chirality flip reactions ($y_{\mbox{\tiny{Yukawa}}}^{\mu}=y_{R_2}-y_{L_2}+y_0$) and extremely small for the weak sphalerons ($y_E$) as compared to that of the electron chirality flip processes ($y_{\mbox{\tiny{Yukawa}}}^e=y_{R_1}-y_{L_1}+y_0$) in the whole interval. As an example, $y_{\mbox{\tiny{Yukawa}}}^{\mu}(1)\simeq 7.71\times10^{-9}$ and $y_E(1)\simeq 5.45\times10^{-13}$, which are negligible as compared to $y_{\mbox{\tiny{Yukawa}}}^e(1)\simeq 0.0066$. 
The value of the fermion chiral asymmetry at $T=T_{EW}$ which is the onset of the electroweak phase transition, can now be estimated as follows: $\sum_f y_{f_R}(1)-y_{f_L}(1)\simeq\sum_i y_{R_i}(1)-y_{L_i}(1)\simeq y_{\mbox{\tiny{Yukawa}}}^e(1)-3y_0(1)\simeq 0.0066-3(0.0002)\simeq6\times10^{-3}$. It should be noted that the contribution of up-type quarks to this asymmetry cancels that of the down-type quarks; therefore, only the contribution of the leptons is taken into account. This chiral asymmetry is important since its evolution is strongly coupled to that of the Maxwellian magnetic fields in the broken phase \cite{Boyarsky}.


As mentioned above, for $k=0.02$  and $c_{\Gamma_1}=0.02$, the minimum value of $\log(y_{R_1}^{(0)})$ which gives the BAU at $T=T_{EW}$, is $\log(y_{R_1}^{(0)})_{min}\simeq 3.45$. We repeat the same investigation with different values of $c_{\Gamma_1}$, then obtain the corresponding $\log(y_{R_1}^{(0)})_{min}$ and present the results in Table \ref{table:k0.02}. It can be seen that $\log(y_{R_1}^{(0)})_{min}$ depends on the chirality flip rate of the electrons and decreases, as the rate is decreased. However, the values of $\log(y_{R_1}^{(0)})_{min}\simeq 5.80\ \textrm{and}\ 4.05$ appearing in the first and second rows of Table 2, are unacceptable, since $\log(y_f)<4$ or equivalently $\frac{\mu_f}{T}<1$ in our model.
Interestingly, we find that as $\log(y_{R_1}^{(0)})$ exceeds $\log(y_{R_1}^{(0)})_{min}$ in each case, ${\eta_B(1)}$ saturates to $\simeq 6.54\times 10^{-10}$. 


We also repeat the above investigations with different values of $k$ and observe almost the same behavior. However, the important and interesting point is that the saturated amount of $\eta_B(1)$ depends solely on the value of $k$ as presented in Table \ref{table:k}. Indeed, the value of $k$ which can lead to the BAU at $T=T_{EW}$ is $k\simeq0.02$.
\begin{table}[ht]
\caption{$B_Y^{(0)}=10^{-2}$G and $k=0.02$}
\label{table:k0.02}
\centering
\begin{tabular}{c c c c c}
\hline 
& $\log(y_{R_1}^{(0)})_{min}$ & $c_{\Gamma_1}$ & $\eta_B(1)$ & $B_Y(1)$ \\[0.5ex]
\hline
& 5.80 & 1 & $6.24\times10^{-10}$ & $1.77\times10^{20}$G \\
& 4.05 & 0.1 & $6.44\times10^{-10}$ & $1.10\times10^{20}$G \\
& 3.70 & 0.05 & $6.48\times10^{-10}$ & $1.08\times10^{20}$G \\
& 3.45 & 0.02 & $6.50\times10^{-10}$ & $1.13\times10^{20}$G \\
& 3.35 & 0.01 & $6.51\times10^{-10}$ & $1.12\times10^{20}$G \\
& 3.25 & 0 & $6.59\times10^{-10}$ & $1.13\times10^{20}$G \\[1ex]
\hline
\end{tabular}
\end{table}

\begin{table}[ht]
\caption{$B_Y^{(0)}=10^{-2}$G}
\label{table:k}
\centering
\begin{tabular}{c c c c c}
\hline 
& $k$ & ${\eta_B(1)}_{sat}$\\[0.5ex]
\hline
& 1 & $3.27\times10^{-8}$ \\
& 0.2 & $6.54\times10^{-9}$ \\
& 0.1 & $3.27\times10^{-9}$ \\
& 0.02 & $6.54\times10^{-10}$ \\
& 0.01 & $3.27\times10^{-10}$ \\[1ex]
\hline
\end{tabular}
\end{table}

\subsection{Production of matter asymmetries by hypermagnetic fields}\label{Production of matter asymmetries by hypermagnetic fields}
As stated earlier, in all parts of this study we have the assumption of $c_1=c_2=c_3=0$. In this subsection we have the extra assumption of zero initial value for the right-handed electron asymmetry which leads to zero initial matter asymmetries.  
Here, we want to investigate whether strong hypermagnetic fields are able to produce and grow matter asymmetries in the presence of the weak sphalerons which try to wash out the asymmetries. As stated in Section \ref{Equilibrium Conditions} and Subsection \ref{Nonequilibrium and Almost Equilibrium Processes}, the initial conditions and evolutions of all of the matter asymmetries are interconnected by the constraints on the system. In particular, the initial hypermagnetic field produces the asymmetries of all of the species simultaneously, including that of the muons.
As stated in the Introduction, the weak sphalerons act only on the left-handed fermions; therefore, the washout process is completed when the Yukawa interactions are also in thermal equilibrium \cite{Harvey}. Indeed, the Yukawa processes are in equlibrium in these high temperatures in the absence of the hypermagnetic fields; however, they can fall out of equilibrium in the presence of the strong hypermagnetic fields. Therefore, there is a possibility for the hypermagnetic fields to produce and grow the matter asymmeties on one hand, and on the other hand protect them from washout by making the Yukawa processes fall out of equlibrium. 
We have hypothesized that the amount of falling out of equlibrium is tiny for fast processes and decided to study the behaviour of the system by considering one extra dynamical Yukawa process other than the electron one to check our hypothesis. Among the Yukawa processes the muon one which is the slowest one after the electron Yukawa process was chosen.

In this subsection, we investigate the possibility to generate and grow matter asymmetries by initial hypermagnetic fields, and especially to obtain not only the final baryonic asymmetry as large as the BAU $\sim 6\times  10^{-10}$, but also a final hypermagnetic field amplitude of the order of $\sim 10^{20}$G at the onset of the EWPT.

\begin{figure} 
  \includegraphics[width=65mm]{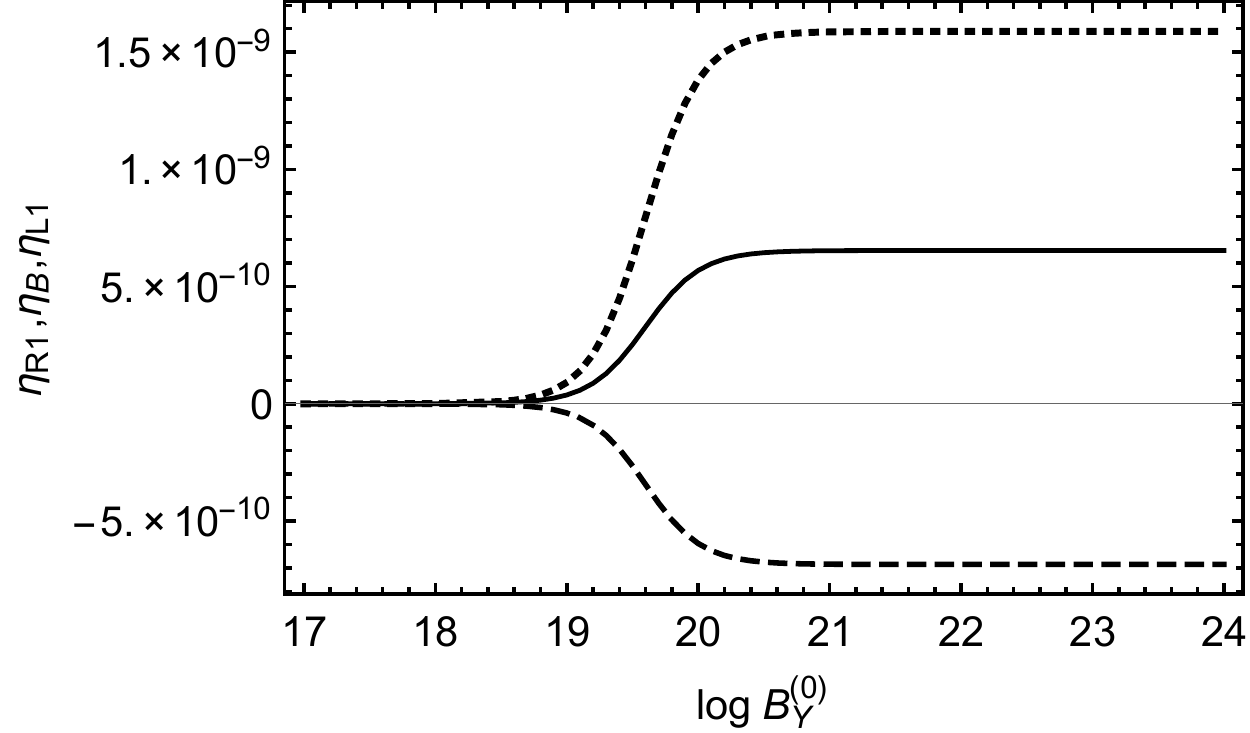}
  \hspace{2mm}
  \includegraphics[width=65mm]{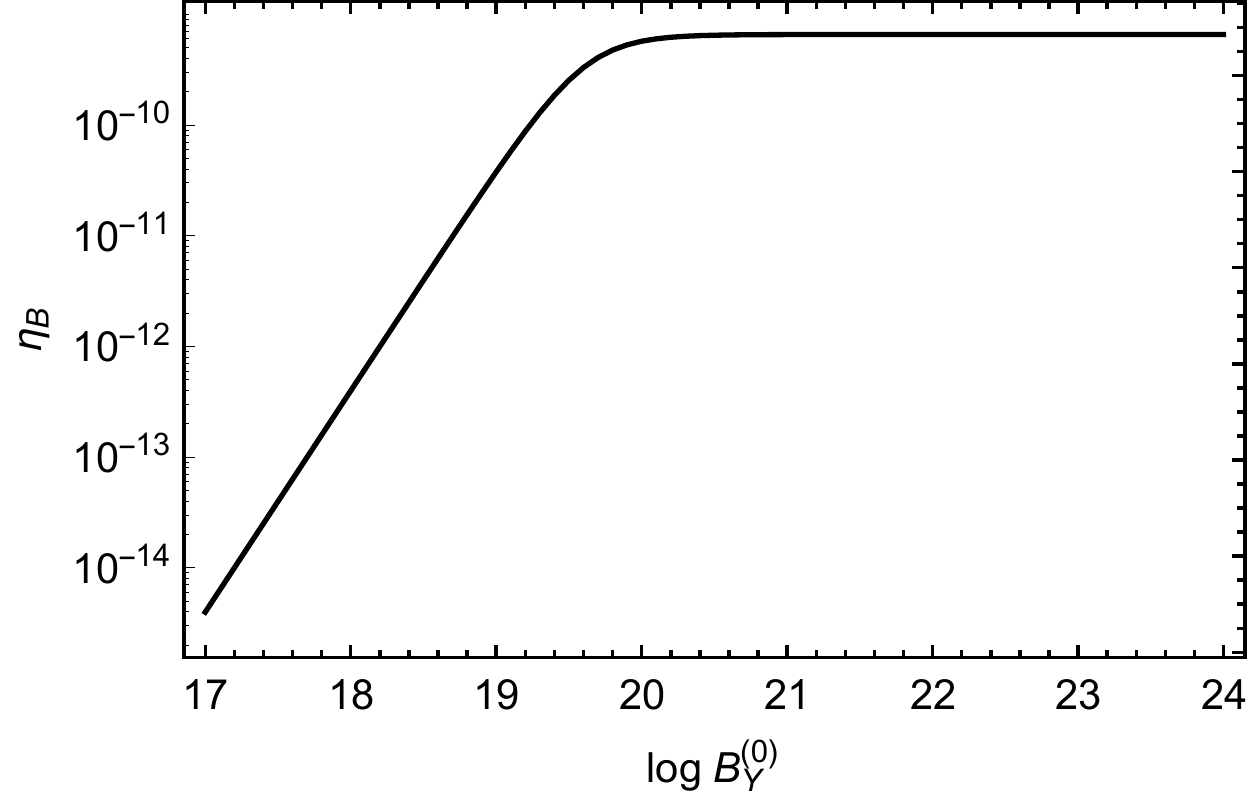}
  \begin{center}
  \includegraphics[width=65mm]{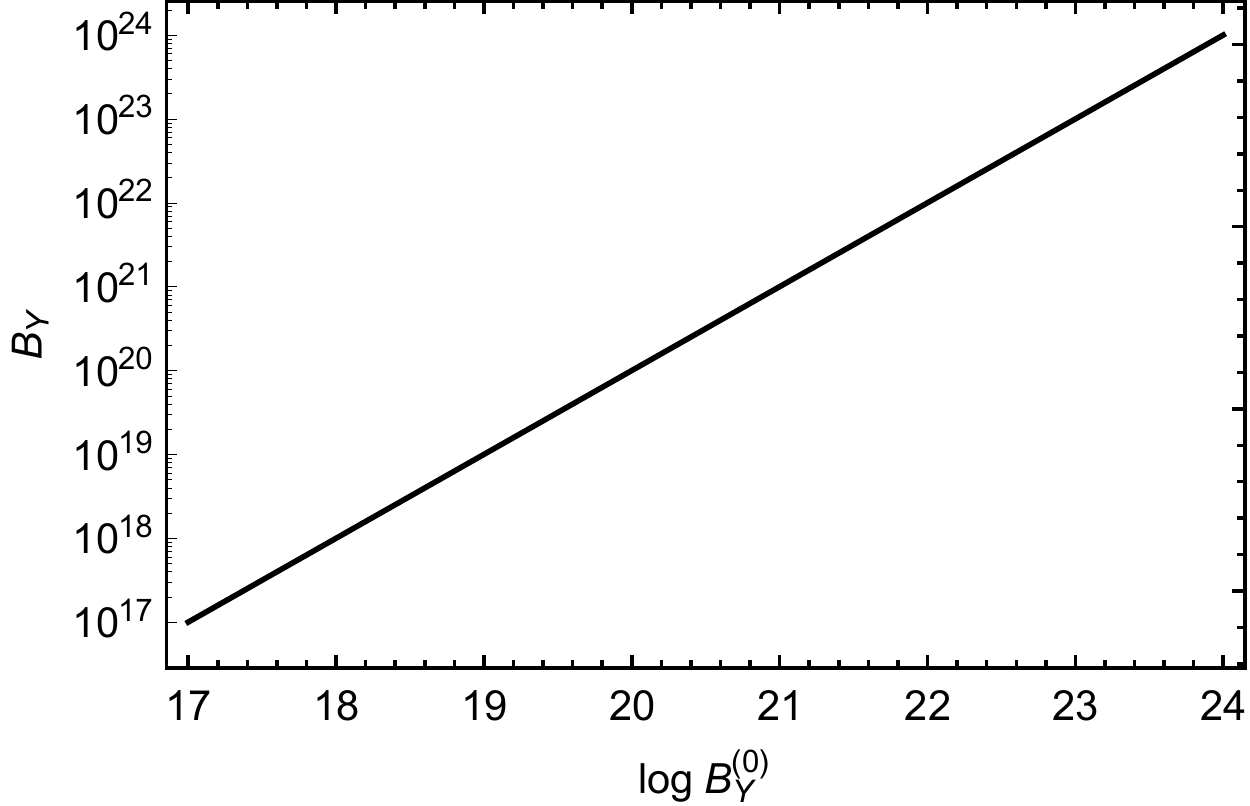}
  \end{center}
  \caption{Top left: The final asymmetries of right-handed electrons $\eta_{R_1}$ (dotted line), baryons $\eta_B$ (solid line), and left-handed electrons $\eta_{L_1}$ (dashed line) at the EWPT time $t_{EW}$. Top right: Log-Log plot for the final asymmetry of baryons $\eta_B$, versus $B_Y^{(0)}$. Bottom: The final amplitude of hypermagnetic field $B_Y$ at the EWPT time $t_{EW}$. It is assumed that all initial asymmetries are zero, $k=0.02$ ($k_0=2\times 10^{-9}T_{EW}$), $c_{\Gamma_1}=1$, and $B_Y^{(0)}$ varies from $10^{17}$G to $10^{24}$G. The maximum relative error for these graphs is of the order of $10^{-16}$.
}\label{one-}
\end{figure}

We solve the evolution equations with zero initial matter asymmetries, $k=0.02$, $c_{\Gamma_1}=1$, and the initial hypermagnetic field amplitude in the range $10^{17}\mbox{G} \leq B_Y^{(0)} \leq10^{24}\mbox{G}$, then obtain the final values of baryonic asymmetry $\eta_B(1)$ and the hypermagnetic field amplitude $B_Y(1)$, and present the results in Figure \ref{one-} and Table \ref{table:c1}. It can be seen that, for $B_Y^{(0)}\lesssim10^{19.5}$G, the growth of $\eta_B(1)$ with respect to $B_Y^{(0)}$ is quadratic but it saturates to $\sim6.54\times10^{-10}$ for $B_Y^{(0)}\gtrsim10^{21}$G. Although $B_Y(1)/B_Y^{(0)}\simeq1$ in the whole range, it increases with a very small rate as $B_Y^{(0)}$ increases, then exceeds 1 at $B_Y^{(0)}\sim10^{21}$G and saturates to $\sim 1.0000047$ for greater values of $B_Y^{(0)}$. Indeed, the minimum value of $B_Y^{(0)}$ which results in the maximum absolute values for the matter asymmetries at $T=T_{EW}$ is, $B_Y^{(0)}\simeq 10^{21}$G. The final values of $\eta_B(1) \simeq 6.5\times10^{-10}$ and $B_Y(1)\simeq 10^{21}$G are obtained for $B_Y^{(0)}\simeq 10^{21}$G and the aforementioned initial conditions, as shown in Table \ref{table:c1}.

\begin{table}[ht]
\caption{$y_{R_1}^{(0)}=0$, $k=0.02$ and $c_{\Gamma_1}$=1}
\label{table:c1}
\centering
\begin{tabular}{c c c c c}
\hline 
& $B_Y^{(0)}$ & $\eta_B(1)$ & $B_Y(1)/B_Y^{(0)}$\\[0.5ex]
\hline
& $10^{18}$G & $3.95\times10^{-13}$ & $0.9999067$ \\
& $10^{18.5}$G & $3.93\times10^{-12}$ & $0.9999069$ \\
& $10^{19}$G & $3.75\times10^{-11}$ & $0.9999091$ \\
& $10^{19.5}$G & $2.53\times10^{-10}$ & $0.9999253$ \\
& $10^{20}$G & $5.69\times10^{-10}$ & $0.9999677$ \\
& $10^{20.5}$G & $6.44\times10^{-10}$ & $0.9999946$ \\
& $10^{21}$G & $6.53\times10^{-10}$ & $1.0000024$ \\
& $10^{21.5}$G & $6.54\times10^{-10}$ & $1.0000042$ \\
& $10^{22}$G & $6.54\times10^{-10}$ & $1.0000046$ \\
& $10^{22.5}$G & $6.54\times10^{-10}$ & $1.0000047$ \\
& $10^{23}$G & $6.54\times10^{-10}$ & $1.0000047$ \\[1ex]
\hline
\end{tabular}
\end{table}

\begin{figure} 
  \includegraphics[width=65mm]{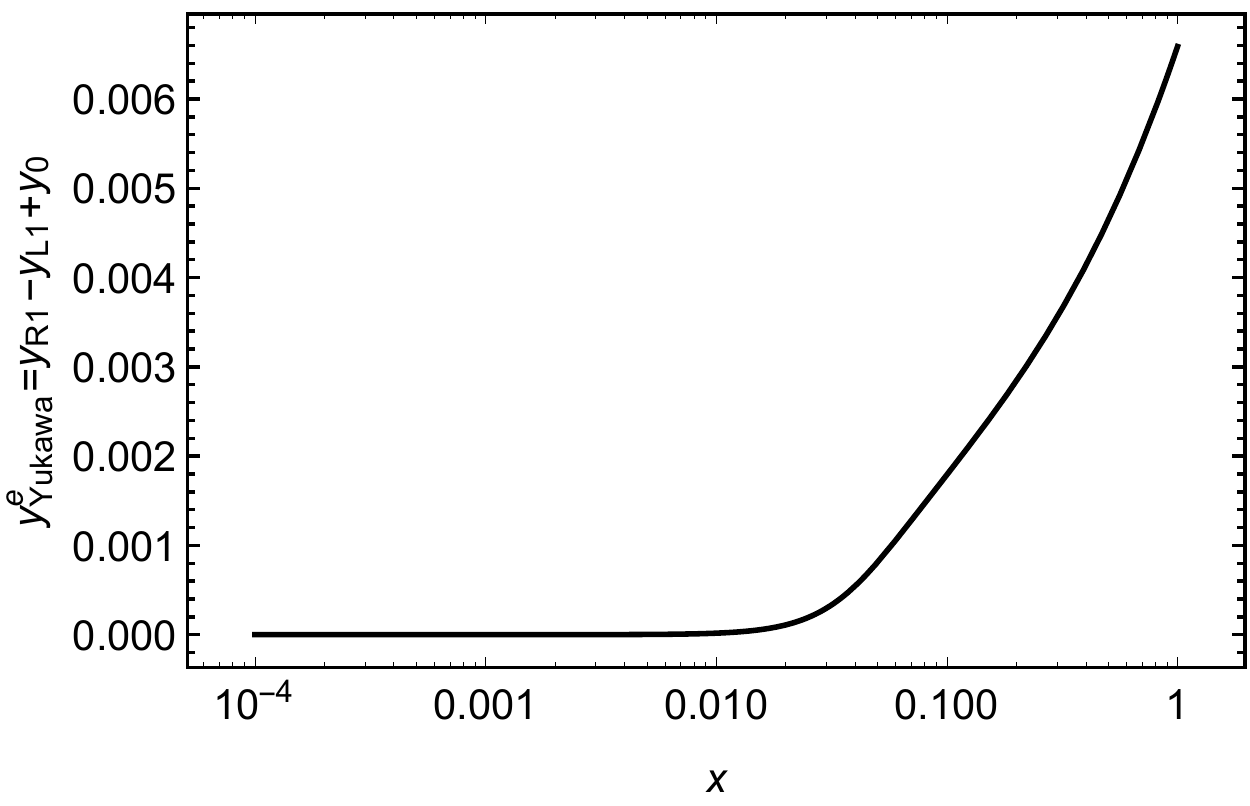}
  \hspace{2mm}
  \includegraphics[width=65mm]{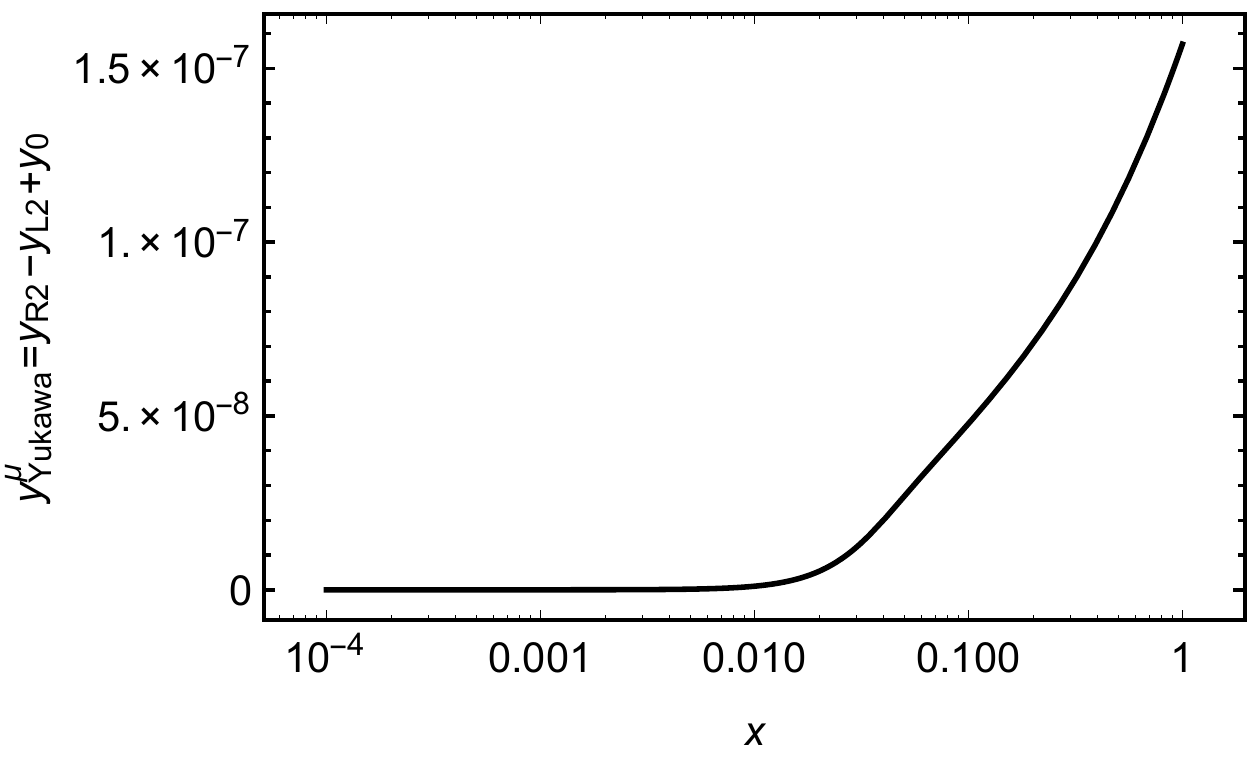}
\begin{center}
  \includegraphics[width=65mm]{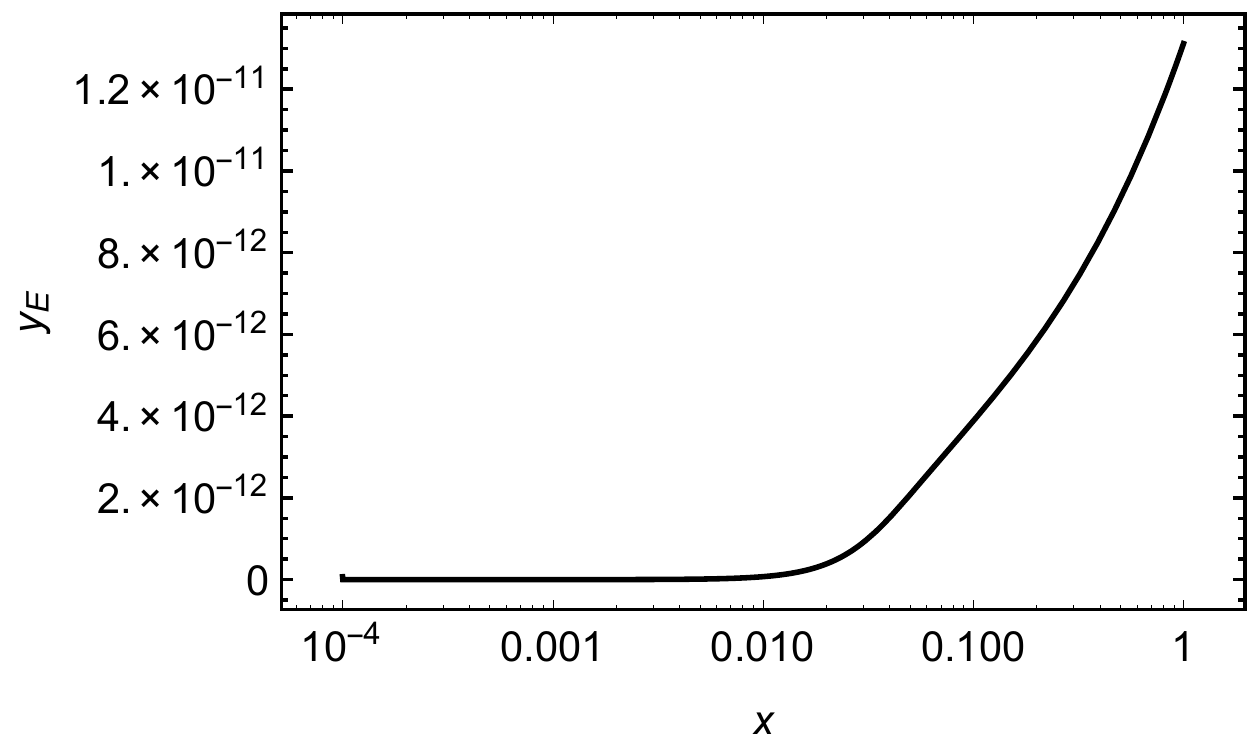}
\end{center}
\includegraphics[width=65mm]{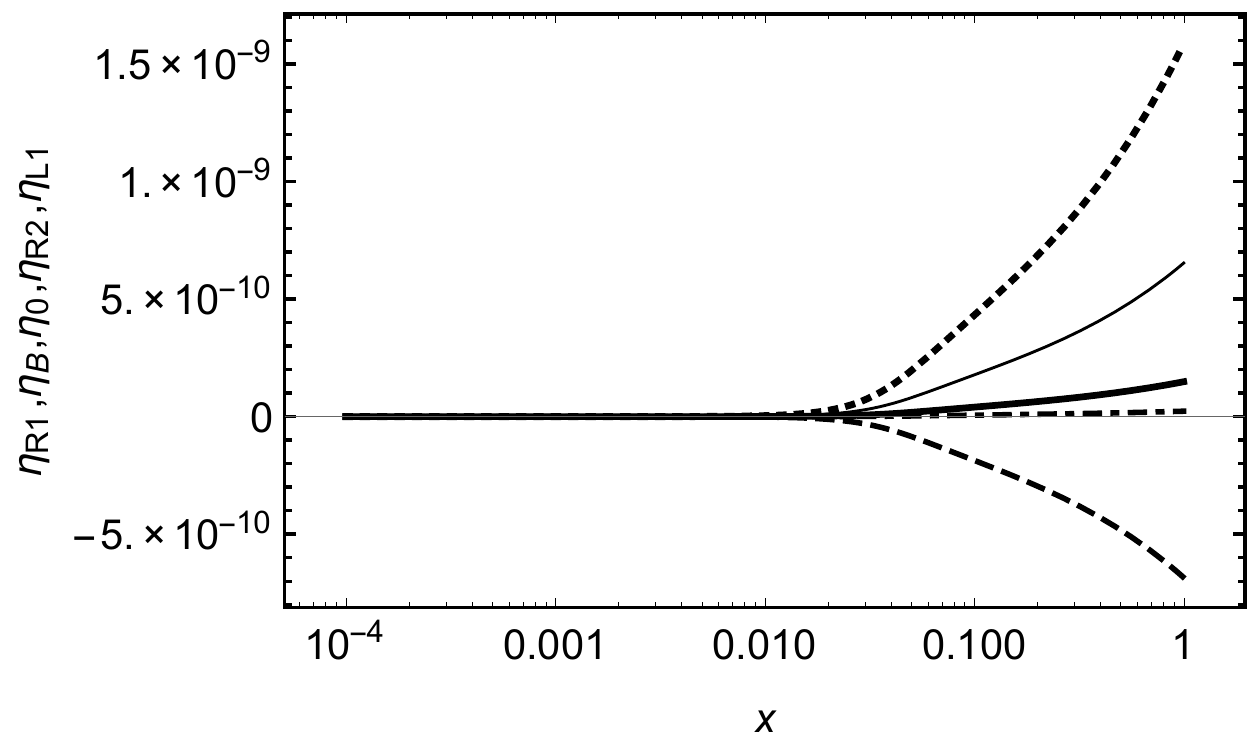}
  \hspace{2mm}
  \includegraphics[width=65mm]{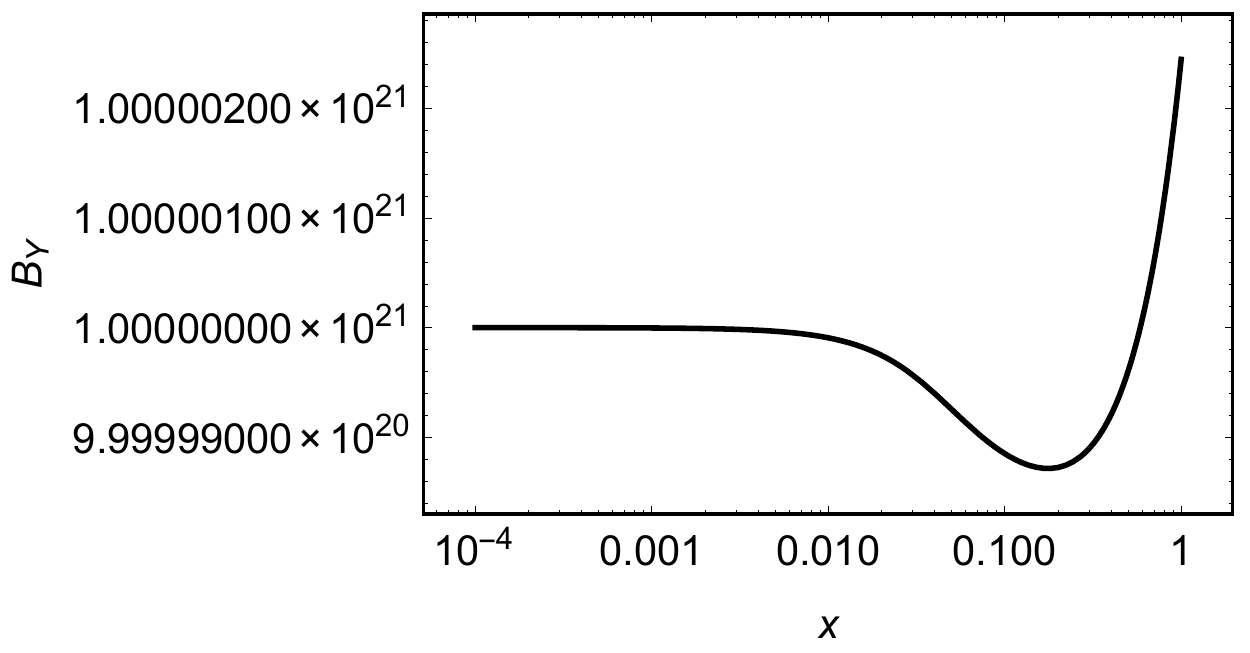}
\caption{The time plots for (top left): $y_{\mbox{\tiny{Yukawa}}}^e=y_{R_1}-y_{L_1}+y_0$, (top right): $y_{\mbox{\tiny{Yukawa}}}^{\mu}=y_{R_2}-y_{L_2}+y_0$, and (middle): $y_E$, representing the amounts of falling out of chemical equilibrium for the electron Yukawa reactions, the muon Yukawa reactions and the weak sphaleron processes, respectively.  The time plots for (bottom left): the asymmetries of right-handed electrons $\eta_{R_1}$ (dotted line), baryons $\eta_B$ (thin line), Higgs bosons $\eta_0$ (thick line), right-handed muons $\eta_{R_2}$ (dotdashed line), left-handed electrons $\eta_{L_1}$ (dashed line), and 
(bottom right): the hypermagnetic field amplitude $B_Y$. The relevant parameters and initial conditions are: $k=0.02$ ($k_0=2\times10^{-9}T_{EW}$), $c_{\Gamma_1}=1$, $B_Y^{(0)}=10^{21}$G, and zero initial matter asymmetries. The maximum relative error for these graphs is of the order of $10^{-17}$.\\
}\label{21-}
\end{figure}

The time plots corresponding to $B_Y^{(0)}=10^{21}$G are shown in Figure \ref{21-}. The bottom plots show the evolution of the hypermagnetic field amplitude and some of the asymmetries. It can be seen that the system goes out of equilibrium for $x\gtrsim 0.01$. The top and middle plots show that, as expected, the amount of falling out of chemical equilibrium is very small for the muon chirality flip reactions and extremely small for the weak sphalerons, as compared to that of the electron chirality flip processes. As an example, $y_{\mbox{\tiny{Yukawa}}}^{\mu}(1)\simeq 1.57\times10^{-7}$ and $y_E(1)\simeq 1.31\times10^{-11}$, which are negligible as compared to $y_{\mbox{\tiny{Yukawa}}}^e(1)\simeq 0.0066$. It is interesting to note that not only $y_{\mbox{\tiny{Yukawa}}}^e(1)$ but also $y_0(1)$ are the same as the ones obtained for the specific case of $\log(y_{R_1}^{(0)})=3.45$ discussed in subsection \ref{Hypermagnetic field growth via matter asymmetries}.
Therefore, 
the same initial value for the fermion chiral asymmetry in the broken phase, $\sum_f \eta_{f_R}(1)-\eta_{f_L}(1)\simeq6\times 10^{-3}$ is obtained as well.

As mentioned above, for $k=0.02$ and $c_{\Gamma_1}=1$, the minimum value of $B_Y^{(0)}$ which gives the BAU at $T=T_{EW}$, is ${B_Y^{(0)}}_{min}\sim10^{21}$G. We repeat the same investigation with different values of $c_{\Gamma_1}$, then obtain the corresponding ${B_Y^{(0)}}_{min}$ and present the results in Table \ref{table:2k0.02}. It can be seen that ${B_Y^{(0)}}_{min}$ depends on the chirality flip rate of the electrons and decreases, as the rate is decreased. However, the most important and interesting point is that, as ${B_Y^{(0)}}$ exceeds ${B_Y^{(0)}}_{min}$ in each case, ${\eta_B(1)}$ and $B_Y(1)/B_Y^{(0)}$ saturate to $\simeq 6.54\times 10^{-10}$ and $\simeq1.0000047$, respectively. It seems that these values are independent of $c_{\Gamma_1}$ and depend solely on the value of $k$. Therefore, we repeat the above investigations with different values of $k$ and present the saturated amounts of $\eta_B(1)$ and $(B_Y(1)/B_Y^{(0)})-1$ for each value of $k$ in Table \ref{table:2k}. Interestingly, the second column of Table \ref{table:2k} matches exactly that of Table \ref{table:k} and again, the value of $k$ which can lead to the BAU at $T=T_{EW} $ is $k\simeq0.02$. More importantly, ${\eta_B(1)}_{sat}$ and $({B_Y(1)/B_Y^{(0)}})_{sat}-1$ are proportional to $k$ and $k^2$, respectively.

\begin{table}[ht]
\caption{$y_{R_1}^{(0)}=0$ and $k=0.02$}
\label{table:2k0.02}
\centering
\begin{tabular}{c c c c c}
\hline 
& ${B_Y^{(0)}}_{min}$ & $c_{\Gamma_1}$ & $\eta_B(1)$ & $B_Y(1)/B_Y^{(0)}$ \\[0.5ex]
\hline
& $10^{21.9}$G & 100 & $6.54\times10^{-10}$ & $1.0000019$ \\
& $10^{21.4}$G & 10 & $6.54\times10^{-10}$ & $1.0000019$ \\
& $10^{21}$G & 1 & $6.54\times10^{-10}$  & $1.0000024$ \\
& $10^{20.5}$G & 0.1 & $6.54\times10^{-10}$  & $1.0000008$ \\
& $10^{20.3}$G & 0.01 & $6.54\times10^{-10}$  & $0.9999994$ \\[1ex]
\hline
\end{tabular}
\end{table}

\begin{table}[ht]
\caption{$y_{R_1}^{(0)}=0$}
\label{table:2k}
\centering
\begin{tabular}{c c c c c}
\hline 
& $k$ & ${\eta_B(1)}_{sat}$ & $({B_Y(1)/B_Y^{(0)}})_{sat}-1$\\[0.5ex]
\hline
& 1 & $3.27\times10^{-8}$ & $1.17\times10^{-2}$\\
& 0.2 & $6.54\times10^{-9}$ & $4.67\times10^{-4}$\\
& 0.1 & $3.27\times10^{-9}$ & $1.16\times10^{-4}$\\
& 0.02 & $6.54\times 10^{-10}$ & $4.67\times10^{-6}$\\
& 0.01 & $3.27\times10^{-10}$ & $1.16\times10^{-6}$\\[1ex]
\hline
\end{tabular}
\end{table}


\newpage

\section{Summary and Discussion}\label{Summary and Discussion}
In this paper we have studied a minimal model for investigating the simultaneous evolution of the matter asymmetries and the hypermagnetic field in the temperature range $T_{EW}<T<T_{RL}$. We have categorized the major reactions as either ``Fast", ``Dynamical" or ``Slow" in the temperature range under study and for $T>T_{RL}$. We have used the Fast processes for $T \geq T_{RL}$ to put constraints on the initial conditions at $T=T_{RL}$, and used those for $T_{EW}<T<T_{RL}$ to reduce the number of dynamical equations. We have included the chemical potentials of all matter fields, including the Higgs field, in our evolution equations. We have shown that in our minimal model all of the conservation laws and constraints are built in and are maintained.

An important part of our study has been to check whether it is possible for the hypermagnetic field to protect the baryonic asymmetry from being washed out by the weak sphalerons. Since the weak sphalerons act only on the left-handed fermions, the washout process can be completed only when the Yukawa interactions are also in equilibrium \cite{Harvey}. However, the strong hypermagnetic field, either present from the beginning or produced by the large initial matter asymmetries, plays 
an important role in this regard. Indeed, it makes 
the system fall out of chemical equilibrium, especially near the EWPT. As a result, it does not allow the washout process be completed and prevents the erasure of the asymmetries by keeping the processes
out of equilibrium. Albeit, the amount of falling out of equilibrium for each process depends on its rate; that is, the smaller the rate, the larger the aforementioned amount. 
Therefore, the amount of falling out of equilibrium for the electron Yukawa reaction is extremely larger than that of the other processes since its rate is very small due to the tiny Yukawa coupling of the electrons. 

The competition between the hypermagnetic field and the weak sphalerons, as descrbed above, has been investigated in detail for two different cases; namely,  hypermagnetic field growth via matter asymmetries, and matter asymmetry generation by hypermagnetic fields, in Subsections \ref{Hypermagnetic field growth via matter asymmetries} and \ref{Production of matter asymmetries by hypermagnetic fields}.
Table \ref{table:2k0.02} shows that the hypermagnetic field is victorious over the weak sphalerons when its initial amlitude is $B_Y^{(0)}\sim 10^{21}$G and the rate of the electron Yukawa reaction is the one estimated in Ref.\ \cite{Kamada} ($c_{\Gamma_1}=1$). In this case, the initial strong hypermagnetic field not only produces and grows the matter asymmetries but also preserves them from washout by the weak sphalerons. The same is true when the estimated rate becomes 10 or 100 times larger ($c_{\Gamma_1}=10$ or $100$); albeit, the initial hypermagnetic field should become stronger as well. Table \ref{table:k0.02} shows that, even for the rate of the electron Yukawa reaction 50 times smaller than the value estimated in Ref.\ \cite{Kamada} ($c_{\Gamma_1}=0.02$), large initial matter asymmetries  ($\log(y_{R_1}^{(0)})_{min}=3.45$) are needed in order to produce the strong hypermagnetic field which can protect the matter asymmetries from washout by the weak sphalerons. When the aforementioned rate becomes larger ($c_{\Gamma_1}=0.05$), the hypermagnetic field will overcome the weak sphalerons if the initial matter asymmetries become larger ($\log(y_{R_1}^{(0)})_{min}=3.7$) as well.

Another important result is that, in both cases studied in Subsections \ref{Hypermagnetic field growth via matter asymmetries} and \ref{Production of matter asymmetries by hypermagnetic fields}, 
the final baryon asymmetry $\eta_B(1)\sim6.5\times10^{-10}$ and the final amplitude of the hypermagnetc field $B_Y(1)\sim 10^{20-21}$G at the EWPT can be obtained by choosing $k\sim0.02$ ($k_0=2\times10^{-9}T_{EW}$) in this model. As mentioned in Section \ref{Introduction}, the amount of BAU as extracted from the observations of CMB or from the abundances of light elements in the IGM is $\eta_B\sim 6\times10^{-10}$. Let us also briefly state some features of the present day magnetic fields obtained from the observations of CMB and gamma rays from blazars. We then check the compatibility of our main results with these observational data.  

The observations of the CMB temperature anisotropy put an upper bound on the strength $B_0$ of the present magnetic fields, $B_0 \lesssim 10^{-9}$G on the CMB scales $\lambda_0\gtrsim 1$Mpc \cite{Ade}. Furthermore, the observations of the gamma rays from blazars not only provide both lower and upper bounds on the strength $B_0$, but also indicate the existence of the large scale magnetic fields with the scales as large as $\lambda_0\simeq1$Mpc \cite{Ando,Essey,Chen}. The strength $B_0$ of the present intergalactic magnetic fields (IGMFs) reported in \cite{Ando} is $B_0\simeq10^{-15}$G. Two different cases are also investigated in Ref.\ \cite{Essey}. In the first case, where blazars are assumed to produce both gamma rays and cosmic rays, they find $1\times 10^{-17}\textrm{G}<B_0<3\times10^{-14}\textrm{G}$. However, in the second case where the cosmic ray component is excluded, they report that the $10^{-17}$G lower limit remains valid but the upper limit depends on the spectral properties of the source. Reference \cite{Chen} estimates the strength of the IGMFs to be in the range $B_0\simeq 10^{-17}-10^{-15}$G, which is consistent with the above mentioned results of \cite{Ando,Essey}. Moreover, a nonvanishing helicity of the present large scale magnetic fields is also infered with the strength $B_0\simeq5.5\times 10^{-14}$G in Ref.\ \cite{Chen2}.

The time evolution of the cosmic magnetic fields can be influenced by various effects, such as the cosmic expansion, the interaction with turbulent fluid (the inverse cascade mechanism), the viscous diffusion, and the Abelian anomalous effects. In the trivial adiabatic evolution of the cosmic magnetic fields due to the cosmic expansion, the strength $B(t)\propto a^{-2}(t)$ and the scale $\lambda(t)\propto a(t)$, where $a(t)$ is the scale factor. It is believed that the plasma becomes neutral after the recombination and therefore, to a good approximation, the magnetic fields evolve trivially \cite{Kamada}. 

Various studies for a maximally helical magnetic field evolving in a turbulent plasma show the approximate conservation of magnetic helicity but the transfer of power from small scales to larger ones \cite{Kahniashvili} due to an inverse cascade mechanism. In this mechanism, which needs large amounts of magnetic helicity to operate correctly \cite{Dvornikov}, $\lambda(t)$ grows faster than $a(t)$ \cite{Fujita} and the spectrum develops with a characteristic scaling law \cite{Campanelli}. Using the scaling relation, the spectrum of the primordial magnetic fields can be expressed in terms of $\lambda_0$ and $B_0$ of present magnetic fields in the following form (see Ref.\ \cite{Fujita} and Appendix C of Ref.\ \cite{Kamada})
\be\begin{split}\label{lambda-B}
B(T)\simeq(1\times10^{20}\textrm{G})(\frac{T}{100\textrm{GeV}})^{7/3}(\frac{B_0}{10^{-14}\textrm{G}})g_B(T),\cr
\lambda(T)\simeq(2\times10^{-29}\textrm{Mpc})(\frac{T}{100\textrm{GeV}})^{-5/3}(\frac{\lambda_0}{1\textrm{pc}})g_{\lambda}(T) 
\end{split}\ee
where $g_B(T)$ and $g_{\lambda}(T)$ are O(1) factors depending on the number of relativistic species. Moreover, assuming that the present magnetic fields have experienced the inverse cascade process, one obtains 
\be\label{lambda0B0}
\frac{\lambda_0}{1\textrm{pc}}\simeq a\frac{B_0}{10^{-14}\textrm{G}},
\ee 
where the constant of proportionality a is model-dependent \cite{Banerjee,Durrer}. Now, we check the compatibility of our results with the observations. To do that, we use the above equations 
to estimate the present scale and amplitude of the magnetic fields resulted from the evolution of the hypermagnetic fields of our model after the EWPT. 


We estimate the scale of the hypermagnetic field used in our investigations by using the relation $\lambda=k^{-1}$, 
and obtain $\lambda(T_{EW})=(2\times10^{-9}T_{EW})^{-1}=3.225\times10^{-26}\mbox{pc}$. Assuming that the time evolution of the magnetic fields from $T=T_{EW}$ till now ($T_0\simeq 2\textrm{K}\simeq17.2\times10^{-14}\mbox{GeV}$) is trivial, we use $\lambda(t) \propto a(t) \propto T^{-1}$ and obtain the present scale of the magnetic fields as $\lambda(T_0)\simeq1.875\times10^{-11}\mbox{pc}.$
Since, the acceptable scales of present magnetic fields are much higher than this value, we should rely on an inverse cascade mechanism. We assume that the only nontrivial process is the inverse cascade process starting immediately after $T=T_{EW}$. We then use Eqs.\ (\ref{lambda-B}), and roughly estimate $\lambda_0$ and $B_0$ for $\lambda(T_{EW})\simeq3.225\times10^{-26}\mbox{pc}$ and $B(T_{EW})\simeq 10^{20}\mbox{G}$ to obtain $\lambda_0\simeq 1.6125\times 10^{-3}\mbox{pc}$ and $B_0\simeq 10^{-14}\mbox{G}$. 
Therefore, in this model which includes the weak sphalerons, the values of the baryonic asymmetry and the amplitude of the magnetic fields are consistent with the current data; however, the scale of the magnetic fields is still much lower than the estimated scales of the magnetic fields in the intergalactic medium. In fact, to the best of our knowledge, there has been no model so far that yields acceptable values for the amplitude and the very large scale of the present magnetic fields at the same time. Since, the evolution of the magnetic fields is affected by so many effects in the history of the Universe, other mechanisms should also be considered to explain this large scale of the magnetic fields. More complex models such as those considering the turbulence driven and anomaly driven inverse cascade mechanisms can also be taken into account both in the symmetric phase and broken phase to enhance the scale of the magnetic fields in future studies (see Refs.\ \cite{Boyarsky, Brandenburg, Rogachevskii, Schober, Vazza}).\\




\noindent Acknowledgements: S. R. would like to thank the School of Particles and Accelerators of IPM. We would like to thank the research office of the Shahid Beheshti University as well.



\end{document}